\documentclass[11pt,a4paper]{article}
\pdfoutput=1

\usepackage{jheppub}
\usepackage[american]{babel}
\def\a{\alpha}
\def\b{\beta}

\def\g{\gamma}

\def\l{\lambda}

\def\o{\omega}

\newcommand{\beq}{\begin{equation}}
\newcommand{\eeq}{\end{equation}}
\newcommand{\bea}{\begin{eqnarray}}
\newcommand{\eea}{\end{eqnarray}}
\newcommand{\nn}{\nonumber}

\allowdisplaybreaks[1]

\title{$AdS_6$ T-duals and Type IIB $AdS_6\times S^2$ Geometries with 7-Branes }

\author[a]{Yolanda Lozano,}
\author[b]{Niall T.~Macpherson,}
\author[a]{Jes\'us Montero}

\affiliation[a]{Department of Physics, University of Oviedo,\\ Avda. Federico Garc\'{\i}a Lorca 18, 33007 Oviedo, Spain}
\affiliation[b]{SISSA International School for Advanced Studies and \\INFN, sezione di Trieste, 34136 Trieste, Italy}

\emailAdd{ylozano@uniovi.es} 
\emailAdd{nmacpher@sissa.it} 
\emailAdd{jesus.montero.a@gmail.com}

\abstract{We show that the first $AdS_6$ backgrounds in Type IIB supergravity known in the literature, namely those constructed via T-duality from the Brandhuber-Oz solution to massive IIA, fit within an extension of the global $AdS_6 \times S^2$ solutions with 7-branes warped over a Riemann surface $\Sigma$, recently classified by D'Hoker, Gutperle and Uhlemann  \cite{DHoker:2017mds,DHoker:2017zwj}, that  describes delocalised 5-branes and 7-branes. The solution constructed through Abelian T-duality provides an explicit example of a Riemann surface with the topology of an annulus, that includes D7/O7-branes. In turn, the solution generated through non-Abelian T-duality arises from the upper half-plane.	
} 

\arxivnumber{1810.08093}

\begin{document}
	
	\maketitle

\section{Introduction}

In recent years there has been remarkable progress in the classification of $AdS_6$ solutions of Type IIB supergravity \cite{DHoker:2017mds,DHoker:2017zwj,D'Hoker:2016rdq,DHoker:2016ysh,Gutperle:2017tjo,Gutperle:2018vdd}. These solutions provide interesting backgrounds in which to realise holographically five dimensional CFTs. Using string theory it was shown  that five dimensional gauge theories, even if non-renomalizable, can be at fixed points in the UV \cite{Seiberg:1996bd,Intriligator:1997pq} (see also \cite{Jefferson:2017ahm}). The holographic study of such fixed point theories is especially relevant, since these theories do not allow a standard Lagrangian description. This direction of study remained however largely unexplored due to the scarcity of explicit $AdS_6$ solutions in supergravity. The only solutions to (massive) Type IIA supergravity, the Brandhuber-Oz (BO) background  \cite{Brandhuber:1999np}, and orbifolds thereof \cite{Bergman:2012kr}, remained the only known solutions\footnote{These solutions contain singularities associated to the presence of a O8 orientifold fixed plane and therefore circumvent the no-go theorem in \cite{Gutowski:2017edr}.} of 10 and 11 dimensional supergravity for many years.  The BO background describes holographically the UV fixed point of a 5d gauge theory with $USp(2N)$ gauge group, an antisymmetric hypermultiplet and $N_f<8$ fundamental hypermultiplets, which can be engineered in string theory on a stack of N D4-branes probing an orientifold 8-plane with $N_f$ D8-branes  \cite{Seiberg:1996bd}\footnote{Orbifolds thereof describe linear quivers that can be engineered on the previous brane system on orbifold singularities \cite{Bergman:2012kr}.}.
In turn, the strong constraints imposed by supersymmetry suggested that no $AdS_6$  solutions existed in Type IIB or M-theory \cite{Passias:2012vp}.

The construction of an explicit $AdS_6$ solution to Type IIB supergravity \cite{Lozano:2012au} by acting with non-Abelian T-duality \cite{Sfetsos:2010uq} on the BO solution, reignited the interest in the study of classifications of $AdS_6$ solutions. It also boosted the applications of non-Abelian T-duality, a transformation known from the 90's at the level of the string worldsheet \cite{delaOssa:1992vci}, as a solution generating technique in supergravity, in particular in the context of Holography \cite{Itsios:2012zv,Itsios:2013wd,Lozano:2011kb,Itsios:2012dc,Barranco:2013fza,Macpherson:2013zba,Jeong:2013jfc,Lozano:2013oma,Gaillard:2013vsa,Elander:2013jqa,Zacarias:2014wta,Caceres:2014uoa,Pradhan:2014zqa,Lozano:2014ata,Sfetsos:2014tza,Kelekci:2014ima,Macpherson:2014eza,Kooner:2014cqa,Araujo:2015npa,Bea:2015fja,Lozano:2015bra,Lozano:2015cra,Araujo:2015dba,Macpherson:2015tka}.
To date many interesting new $AdS$ solutions have been generated using this technique. Some of them evade existing classifications   \cite{Lozano:2012au,Macpherson:2014eza,Bea:2015fja, Lozano:2015bra}, while some others provide non-trivial explicit examples in  certain classes \cite{Lozano:2015bra,Lozano:2015cra}. The interpretation of the transformation at the level of the dual CFT has also been addressed, and an interesting connection with CFTs arising from Hanany-Witten brane set-ups \cite{Hanany:1996ie} involving NS5-branes has been identified \cite{Lozano:2016kum,Lozano:2016wrs,Lozano:2017ole,Itsios:2017cew}.

First attempts at classifying $AdS_6$ solutions were made in \cite{Apruzzi:2014qva,Kim:2015hya,Kim:2016rhs} (see also \cite{Gutowski:2017edr}). These works confirmed the absence of solutions in M-theory \cite{Passias:2012vp} and obtained the BPS equations for Type IIB fields. These equations were tested against the non-Abelian and Abelian T-duals of the BO solution, obtained in \cite{Lozano:2012au} (see also \cite{Lozano:2013oma}).
Later, D'Hoker, Gutperle, Karch and Uhlemann (DGKU) \cite{D'Hoker:2016rdq} derived the reduced BPS equations compatible with the symmetries of $F(4)$, the unique superconformal algebra in 5d \cite{Nahm:1977tg}, and finally obtained the complete local solutions to these equations in terms of two arbitrary holomorphic functions\footnote{See also \cite{Macpherson:2016xwk,Apruzzi:2018cvq} for an alternative formulation, where $AdS_6$ solutions are in one to one correspondence with the solutions of a single Laplace equation and its derivatives.}.  The full non-linear Kaluza-Klein reduction of these Type IIB solutions to 6d $F(4)$ gauged supergravity \cite{Romans:1985tw} has been achieved recently in \cite{Hong:2018amk}. This work generalises the results of \cite{Jeong:2013jfc}\footnote{See also \cite{Malek:2018zcz} for a derivation in the exceptional field theory framework.}, where the consistent truncation of the Type IIB non-Abelian T-dual of the Brandhuber-Oz solution was worked out.

Global physically sensible solutions to the set of equations derived in \cite{D'Hoker:2016rdq}  were constructed by D'Hoker, Gutperle and Uhlemann (DGU) in \cite{DHoker:2017mds}. This work provided the first firm candidates for holographic duals to 5d CFTs living in $(p,q)$ 5-brane 
webs \cite{Aharony:1997ju,Aharony:1997bh}\footnote{These solutions contain singularities associated to the $(p,q)$ 5-branes and therefore circumvent as well the no-go theorem in \cite{Gutowski:2017edr}. This is also the case for the Abelian and non-Abelian T-dual solutions previously discussed.}. These solutions were extended to allow 7-branes in \cite{DHoker:2017zwj}, thus enlarging the classes of dual 5d CFTs to those including this type of branes \cite{DeWolfe:1999hj}.

The interpretation of the geometries constructed in \cite{DHoker:2017mds} as holographic duals of 5d CFTs living in 5-brane webs is by now supported by a variety of tests. The entanglement entropy and free energy of the solutions were computed in \cite{Gutperle:2017tjo}. Recently, the five sphere partition function and conformal central charge have been shown to agree with the corresponding field theory results computed using localisation \cite{Fluder:2018chf}. 
The identification of the bulk states \cite{Bergman:2018hin} and certain spin-2 fluctuations \cite{Passias:2018swc,Gutperle:2018wuk} that are dual to certain classes of operators in the CFT has also been worked out. Some of these tests have been extended to include 7-branes  \cite{Gutperle:2018vdd,Bergman:2018hin}. The behaviour of $(p,q)$-strings has been explored in \cite{Kaidi:2017bmd}. Further support for the holographic interpretation of the solutions in \cite{DHoker:2017mds} has been provided very recently by uplifting the $(p,q)$ 5-brane webs to punctured M5-branes wrapping a Riemann surface, that can be completely specified in terms of the corresponding $AdS_6\times S^2$ geometry \cite{Kaidi:2018zkx}.

In this paper we extend the study of global solutions by DGU and show that the Abelian and non-Abelian T-dual solutions constructed from the BO background fit within this extension.
Our findings show, in particular, that a generalisation of the formalism in  \cite{DHoker:2017mds,DHoker:2017zwj}  allows for a description of delocalised 5 and 7 branes intersections that include orientifold fixed planes.
The solutions constructed in  \cite{DHoker:2017mds} take the form of a warped product of an $AdS_6\times S^2$ geometry over a two dimensional Riemann surface, which needs to have a boundary for regularity. When the Riemann surface is the upper half-plane the solutions contain poles in the real axis that are associated to semi-infinite $(p,q)$ 5-branes. The addition of 7-branes adds punctures which introduce non-trivial monodromies. The solutions constructed via T-duality contain singularities that, rather than being local, correspond to branes smeared along some directions. One such singularity is associated to the location of an orientifold fixed plane, not present in the formalism in  \cite{DHoker:2017mds,DHoker:2017zwj}, related by T-duality to the O8 plane of the BO solution. We show that these solutions fit within an extension of the classification of \cite{DHoker:2017mds,DHoker:2017zwj} in which the Riemann surface has two boundaries that lie infinitely apart from each other. Smeared NS5 branes lie on one of these boundaries, and D7/O7-branes
introduce (smeared) punctures that lie at the other boundary. The Abelian T-dual solution is associated to a Riemann surface with the topology of an annulus, and the one generated through non-Abelian T-duality is associated to the half-plane.

The paper is organised as follows. In section \ref{sec:alaDHoker} we summarise the main features of the local and global $AdS_6$ solutions to Type IIB supergravity constructed in \cite{D'Hoker:2016rdq} and  \cite{DHoker:2017mds,DHoker:2017zwj}. This includes a detailed analysis of the global solutions constructed in \cite{DHoker:2017mds} for the upper half-plane and the annulus, as well as an account of how 7-branes are introduced for the upper half-plane \cite{DHoker:2017zwj}. In section \ref{rs7branesannulus} we work out the generalisation of the annulus to include 7-branes, necessary for the realisation of the Abelian T-dual $AdS_6$ solution as a DGU geometry.
In section \ref{sec:gravity_solutions} we summarise the main features of the Abelian and non-Abelian T-duals of the BO solution, constructed in \cite{Lozano:2012au,Lozano:2013oma}.
In section \ref{Abelian-T-dual} we recover the Abelian T-dual solution from the annulus. We highlight that this is the first explicit solution that has been found with this topology.  In section \ref{sec:NATD_DGU} we show that the non-Abelian T-dual solution arises from an infinite strip whose two boundaries lie infinitely apart from each other. This allows us to use the results for the upper half-plane in \cite{DHoker:2017mds,DHoker:2017zwj}  to recover the solution. 
In section \ref{entanglement} we show that the entanglement entropy associated to the T-dual solutions can be obtained from the general expressions in \cite{Gutperle:2017tjo}. This calculation confirms the CFT dual to the Abelian T-dual solution and hints at a dual CFT based on $USp(2N)$ gauge groups for the non-Abelian T-dual one. Further, in section \ref{field-theory} we discuss a possible CFT interpretation of the non-Abelian T-dual solution. In section \ref{conclusions} we present our conclusions and a summary of open directions related to our study. In Appendix \ref{Annulusappendix} we include further technical details of our computations for the annulus. In Appendix \ref{worldsheet} we use the string sigma model to derive the transformation of worldsheet parity reversal under non-Abelian T-duality, relevant for the identification of the dual of the orientifold 8-plane of the BO solution. Finally, in Appendix \ref{spin2} we briefly review and expand the results of \cite{Gutperle:2018wuk} concerning spin-2 fluctuations of DGKU/DGU geometries to include some excitations around the non-Abelian T-dual solution.

\section{The DGKU/DGU  $AdS_6\times S^2 \times \Sigma$ solutions to Type IIB}\label{sec:alaDHoker}

In this section we summarise the main features of the $AdS_6\times S^2\times \Sigma$ solutions to Type IIB supergravity constructed in \cite{D'Hoker:2016rdq},\cite{DHoker:2017mds}, including as well 7-branes \cite{DHoker:2017zwj}. We review in detail the construction when $\Sigma$ is the upper half-plane, relevant for the realisation of the non-Abelian T-dual solution as a DGU geometry, and of the annulus, relevant for the realisation of the Abelian T-dual solution. In this last case we generalise the annulus construction in \cite{DHoker:2017mds} to include 7-branes. These original results are presented in the separated section \ref{rs7branesannulus}, supplemented by Appendix \ref{Annulusappendix}. 

In \cite{D'Hoker:2016rdq} the problem of finding supersymmetric $AdS_6$ solutions to type IIB supergravity was reduced to finding two holomorphic functions $\mathcal{A}_{\pm}$. Formally, this put things on the same level as the half BPS $AdS_4$ solutions constructed in \cite{D'Hoker:2007xy,D'Hoker:2007xz}.

The local form of the $AdS_6$ solutions in \cite{D'Hoker:2016rdq}  is as follows
\begin{align}\label{eq:metric_DHoker}
ds^2 &= \lambda_6^2 ds^2(AdS_6)+ \lambda_2^2 ds^2(S^2)+ds^2(\Sigma),\quad ds^2(\Sigma)= 4 \tilde{\rho}^2dw d\overline{w} ,\nn\\[2mm]
B_2+i\, C_2 &= \mathcal{C}\wedge \text{Vol}(S^2),
\end{align}
where, $\lambda_6^2,\lambda_2^2, {\tilde \rho}^2, \mathcal{C}$, the dilaton and axion have support in $\Sigma$ only.
Two locally holomorphic functions $\mathcal{A}_{\pm}$ are introduced in terms of which the following can be defined
\begin{align}\label{eq:functions_DHoker}
&\kappa_{\pm}= \partial_w \mathcal{A}_{\pm},\quad \kappa^2= -|\kappa_+|^2+ |\kappa_-|^2,\quad \partial_w \mathcal{B} = \mathcal{A}_+\partial_w\mathcal{A}_--\mathcal{A}_-\partial_w\mathcal{A}_+,\\[3mm] 
&\mathcal{G}= |\mathcal{A}_+|^2-|\mathcal{A}_-|^2+\mathcal{B}+\overline{\mathcal{B}},\quad W = R+\frac{1}{R} = 2 + 6 \frac{\kappa^2\mathcal{G}}{|\partial_w\mathcal{G}|^2},~~ \kappa^2= -\partial_w \partial_{\overline{w}}\mathcal{G}\nonumber .
\end{align}
From these the warp factors of the string frame metric are\footnote{The dilaton appearing here is the usual dilaton, $\Phi=2\phi$ where $\phi$ is the one defined in \cite{D'Hoker:2016rdq}. Note also that Einstein frame metric is used in that paper.}
\beq\label{eq:factors_DHoker}
\lambda_2^2 =e^{\Phi}\frac{ c^2 \kappa^2(1-R)}{9 \tilde{\rho}^2(1+R)},\quad \lambda_6^2= e^{\Phi}\frac{c^2 \kappa^2 (1+R)}{\tilde{\rho}^2(1-R)},~~~\tilde{\rho}^2=  e^{\Phi/2}\frac{\sqrt{R+R^2}}{|\partial_w \mathcal{G}|}\left(\frac{\kappa^2}{1-R}\right)^{3/2}.
\eeq
The axion-dilaton field $B$, such that
\beq\label{eq:axio-dilaton_DHoker_1}
B=\frac{1+i \tau}{1-i \tau},~~~\tau = C_0+ i e^{-\Phi},\qquad F_1= dC_0,
\eeq
is given by
\beq\label{eq:axio-dilaton_DHoker_2}
B= \frac{\kappa_+ \partial_{\overline {w}} \mathcal{G}- \overline{\kappa}_- R \partial_w \mathcal{G}}{\overline{\kappa}_+ R  \partial_w \mathcal{G}- \kappa_-  \partial_{\overline {w}} \mathcal{G}}\, ,
\eeq
and the flux function is
\beq\label{eq:C_potential_DHoker}
\mathcal{C} = \frac{4 i}{9} \bigg(\frac{\overline{\kappa}_- W\partial_w \mathcal{G}-2 \kappa_+ \partial_{\overline{w}}\mathcal{G}}{(W+2)\kappa^2}-\overline{\mathcal{A}}_--2\mathcal{A}_+-\mathcal{K}_0\bigg),
\eeq
where $\mathcal{K}_0$ is an integration constant. 

These solutions preserve 16 supersymmetries, and are invariant under the superalgebra $F(4)$, whose maximal bosonic subalgebra is $SO(2,5) \oplus SO(3)$. 

The existence of physically sensible solutions requires that $\kappa$ and ${\cal G}$ satisfy the conditions $\kappa^2>0$ and ${\cal G}>0$ in the interior of $\Sigma$  \cite{D'Hoker:2016rdq}. This guarantees that $\lambda_6^2$, $\lambda_2^2$ and ${\tilde \rho}^2$ are real and positive. Regular (away from sources) solutions require $\Sigma$ to have a non-empty boundary \cite{DHoker:2017mds}. If we choose to populate it with localised sources, this requires  
\begin{equation}
\label{eq:kappa_G_boundary}
\kappa^2=0 \qquad {\rm and} \qquad {\cal G}=0
\end{equation}
on the boundary\footnote{The first of these needs only hold away from the poles.}.
These conditions guarantee that the 2-sphere shrinks smoothly and that the $AdS_6$ radius is non-vanishing.
The second of these conditions implies that ${\cal G}>0$ in the interior of $\Sigma$ if $\kappa^2 >0$.

The most
general solutions satisfying these regularity conditions for the case in which $\Sigma$ has genus zero and one single boundary were constructed in \cite{DHoker:2017mds}. It was shown that these solutions contain an arbitrary number of asymptotic regions where they behave like the near brane limit of $(p,q)$ 5-branes. More generally though $\Sigma$ can have several boundaries.

Very briefly, the construction in  \cite{DHoker:2017mds} starts by solving the equation $\partial_{\bar w}\partial_w {\cal G}=-\kappa^2$ together with the boundary condition ${\cal G}|_{\partial\Sigma}=0$. This is solved in terms of the scalar Green function on $\Sigma$, $G(w,z)$:
\begin{equation}\label{eq:Green}
{\cal G}(w)=\frac{1}{\pi}\int_\Sigma d^2 z\, G(w,z)\kappa^2(z)
\end{equation}
which satisfies
\beq
\label{Green1}
\partial_{\hat w}\partial_w G(w,z)=-\pi \delta(w,z),~~~~G(w,z){\bigg|}_{w\in \partial\Sigma}=0\, .
\eeq 
The next step is to introduce the meromorphic function $\lambda$ on $\Sigma$,
\begin{equation}\label{eq:meromorphic}
\lambda(w)=\frac{\partial_w {\cal A}_+(w)}{\partial_w {\cal A}_-(w)},~~~~|\lambda|\leq 1 ,
\end{equation}
with $|\lambda|= 1$ only on the boundary. $\lambda$ is then solved using an electrostatics analogy. Namely, the function $-\log{|\lambda|^2}$ is taken to be an electrostatic potential, which is real, locally harmonic and strictly positive in the interior of $\Sigma$, and zero on the boundary, since $\kappa^2=0$ on the boundary implies that $|\partial_w {\cal A}_+|=|\partial_w {\cal A}_-|$.
This potential must then have singularities, which are located at the zeros of $\lambda$. Thus, one finds
\begin{equation}\label{eq:meromorphic_green}
-\log{|\lambda (w)|^2}=\sum_{n=1}^N q_n G(w,s_n)\, ,
\end{equation}
where $G$ vanishes at the boundary due to the second equation in \eqref{Green1}, $s_n$ are the zeros of $\lambda$ in the interior of $\Sigma$ and $q_n$ are integer charges such that $\lambda(w)$ is single-valued.
From here $\lambda(w)$ is obtained by holomorphically splitting this expression, and after some work, it is possible to extract a $\partial_w\cal{A}_{\pm}$ which automatically satisfies the regularity conditions for $\kappa^2$ and can be made to satisfy
\beq\label{eq:dercond}
\partial_{\overline{w}}{\cal A}_{\pm}(\overline{w})=-\overline{\partial_{w}{\cal A}_{\mp}(w)}.
\eeq
From here it remains to integrate $\partial_w\cal{A}_{\pm}$ and $\partial_w\cal{B}$ to construct ${\cal G}(w)$, and then impose ${\cal G}(w)=0$ for $w\in  \partial \Sigma$. From \eqref{eq:dercond} it follows that ${\cal G}$ is piece-wise constant on a boundary provided
\beq\label{eq:peicewiseconstantcond}
\partial_{w}{\cal A}_{\pm}(w)=\partial_{\overline{w}}{\cal A}_{\pm}(\overline{w}),~~~~w\in \partial\Sigma,
\eeq
so when this holds, it is sufficient to impose that ${\cal G}(w)=0$ in any regular region of each boundary, and that there is no monodromy as one moves across the poles in each boundary - this imposes a number of constraints on the loci of the poles, zeros and integration constants which increases with the number of boundaries that enclose the Riemann surface.

\subsection{Realising $(p,q)$ 7-branes}
\label{rs7branes}
Generically, localised 7-branes on a Riemann surface are located on the interior, rather than the boundary. As explained at greater length in  \cite{DHoker:2017zwj}, the 7-branes give rise to $SL(2,\mathbb{R})$ monodromies that leave ${\cal G}$ invariant as one moves around the branch point at the loci of theses branes. This means that the holomorphic functions must transform as
\beq\label{eq:monodromy}
{\cal A_+}\to u_{[p,q]} {\cal A}_+- v_{[p,q]}{\cal A}_- + a,~~~~
{\cal A_-}\to-  \overline{v_{[p,q]}} {\cal A}_+- \overline{u_{[p,q]}}{\cal A}_- + \overline{a},
\eeq
where
\beq\label{eq:uveta}
u_{[p,q]}=\frac{1+\eta_+\eta_-}{2\eta_-},~~~v_{[p,q]}=\frac{1-\eta_+\eta_-}{2\eta_-},~~~~\eta_{\pm}=p\pm i q,~~~~|u_{[p,q]}|^2-|v_{[p,q]}|^2=1,
\eeq
for $(p,q)$ 7-branes.
${\cal G}$ and $\kappa^2$ are invariant under these transformations, and likewise the Einstein frame metric. The axion-dilaton scalar $B$ and flux potential ${\cal C}$ transform in turn as
\beq\label{eq:sl2rtrans}
B\to \frac{ u_{[p,q]} B+ v_{[p,q]}}{\overline{v_{[p,q]}} B+\overline{u_{[p,q]}}},~~~~ {\cal C}\to u_{[p,q]}{\cal C}+v_{[p,q]}\overline{{\cal C}}+{\cal C}_0.
\eeq

In order to realise such an arbitrary number of $(p,q)$ 7-branes at loci $w=w_i$ with commuting monodromies, \cite{DHoker:2017zwj} introduce a function $f$ with the following properties
\begin{align}\label{eq:frules}
-&(f+\overline{f})\geq 0,\nn\\[2mm]
&\overline{f(\overline{w})}=-f(w),\nn\\[2mm]
&f(w)\Big\lvert_{w\sim w_i} \!\!\!=\frac{n_i^2}{4\pi} \log\left(\gamma_i(w-w_i)\right)+O(w-w_i)^0,
\end{align}
where $n_i \in \mathbb{R}$ and equality of the first condition happens only on the boundary of $\Sigma$.  The last condition holds in the neighbourhood of a branch point and implies the following monodromy 
\beq
f(w_i + e^{2\pi i}(w-w_i))= f(w)+\frac{i}{2}n_i^2,
\eeq
when moving around it. The constant $\gamma_i$ is a phase related to the branch cut $w_{b,i}$ associated to $w_i$. These may be parameterised in terms of $c\in[0,1]$ as
\beq
w_{b,i}(c)= w_i+c\frac{w_i-\overline{w_i}}{c-\gamma_i}.
\eeq
One can then define holomorphic derivatives that manifestly give rise to $SL(2,\mathbb{R})$ monodromies due to the properties of $f$, namely
\begin{align}\label{eq:derivativeswith7branes}
\partial_{w}{\cal A}_{+} &= u_{[p,q]} \partial_{w}{\cal A}^{(0)}_+- v_{[p,q]}\partial_{w}{\cal A}^{(0)}_- + \eta_+ f\left(\partial_w {\cal A}^{(0)}_+-\partial_w {\cal A}^{(0)}_-\right),\nn\\[2mm]
\partial_{w}{\cal A}_{-} &= -\overline{v}_{[p,q]} \partial_{w}{\cal A}^{(0)}_++ \overline{u}_{[p,q]}\partial_{w}{\cal A}^{(0)}_- + \eta_- f\left(\partial_w {\cal A}^{(0)}_+-\partial_w {\cal A}^{(0)}_-\right),
\end{align}
where ${\cal A}^{(0)}_\pm$ are the holomorphic functions of some regular seed solution without monodromy that one wishes to add 7-branes to. This set-up has the advantage that
\beq
\kappa^2=(\kappa^{(0)})^2- (f+ \overline{f})\big|\partial_w {\cal A}^{(0)}_+-\partial_w {\cal A}^{(0)}_-\big|^2,
\eeq
which automatically satisfies the regularity conditions for $\kappa^2$.\\

As was the case in the absence of 7-branes, here one then needs to integrate $\partial_w\cal{A}_{\pm}$ and $\partial_w\cal{B}$ and impose that ${\cal G}(w)=0$ on each boundary. This once more constrains the a priori free parameters of a local solution. Generically one can integrate \eqref{eq:derivativeswith7branes} as
\beq
{\cal A}_{\pm}={\cal A}^s_{\pm}+\eta_{\pm}\int^{w}_{w_0}dz \partial_{z}({\cal A}^{(0)}_+(z) -{\cal A}^{(0)}_-(z) )f(z)
\eeq
where $w_0$ is an arbitrary reference point which is assumed to be regular, and ${\cal A}^s_{\pm}$ is the local integral of the single valued part of \eqref{eq:derivativeswith7branes}. Given this, one can integrate $\partial_w {\cal B}$ as
\beq
{\cal B}(w)= {\cal B}^{(0)}(w)+2\int_{w_0}^wdz  {\cal A}^{(0)}(z)\partial_{z}{\cal A}^{(0)}(z) f(z)- {\cal A}^{(0)}(w)\int_{w_0}^w dz \partial_{w}{\cal A}^{(0)}(z) f(z),
\eeq 
where the short hand notation
\beq
{\cal A}^{(0)}(w)={\cal A}^{(0)}_+(w)- {\cal A}^{(0)}_-(w)
\eeq
has been introduced.
${\cal B}^{(0)}$ is formed from ${\cal A}^{(0)}_{\pm}$ only and double integrals are avoided by judiciously integrating by parts\footnote{That ${\cal A}^{(0)}$ appears rather than ${\cal A}^s$ is because the later is an $SL(2,\mathbb{R})$ transformation of the former, under which $\cal{B}$ and ${\cal G}$ are invariant.}. Putting this all together one finds 
\begin{align}\label{eq:calG7brane}
{\cal G}(w)&= {\cal G}^{(0)}(w)+\bigg[2\int_{w_0}^w dz{\cal A}^{(0)}(z)\partial_z{\cal A}^{(0)}(z)f(z)+\text{c.c}\bigg]\nn\\[2mm]
&-\big({\cal A}^{(0)}(z)-\overline{{\cal A}^{(0)}(z)}\big)\bigg[\int_{w_0}^w\partial_z {\cal A}^{(0)}(z)f(z)-\text{c.c}\bigg].
\end{align}

A new ingredient in the presence of the 7-branes is that one must impose by hand that the monodromy when moving round a branch point induces an $SL(2,\mathbb{R})$ transformation of ${\cal A}_{\pm}(w)$ that leaves ${\cal G}$ invariant - ie one of the form \eqref{eq:monodromy}. It is not hard to repackage $\partial_w{\cal A}_{\pm}(w_i+e^{2\pi i}(w-w_i))$ as an $SL(2,\mathbb{R})$ transformation of $\partial_w{\cal A}_{\pm}(w)$ - the only issue is the additional constant shifts that appear in the transformation of ${\cal A}_{\pm}(w)$. Imposing that these are a conjugate pair as in \eqref{eq:monodromy} requires that one fixes
\beq\label{eq: monodomycond}
\text{Re}({\cal A}^{(0)}_+(w_i)-{\cal A}^{(0)}_-(w_i))=\text{Re}(\eta_- {\cal A}^{s}_+(w_i)-\eta_+{\cal A}^{s}_-(w_i))=0,
\eeq
so there is one additional constraint that must be solved for each branch point - here the first equality is true in general, see  \cite{DHoker:2017zwj} for details.

Assuming that one is able to simultaneously solve \eqref{eq: monodomycond} and ${\cal G}=0$ on the boundary,  one is guaranteed to have 7 brane behaviour at $w_i$ by the derivation in section 3.7.1 of  \cite{DHoker:2017zwj}, which does not depend on a specific choice of ${\cal A}^{(0)}_{\pm}$ and $f$.

\subsection{Global solutions for the upper half-plane}
Taking $\Sigma$ to be the upper half-plane, the appropriate Green function is
\beq
G(w,z) = -\log\left|\frac{w-z}{w-\overline{z}}\right|^2,
\eeq
which leads to the meromorphic function
\beq
\lambda(w) = \lambda_0^2 \prod_{n=1}^N\frac{w-s_n}{w-\overline{s}_n},~~~~\text{Im}(s_n)>0,~~~~|\lambda_0|^2=1.
\eeq
Following the details in reference \cite{DHoker:2017mds} one arrives at the following expressions for ${\cal A}_\pm$ 
\begin{equation}
\label{Apm}
{\cal A}_\pm = {\cal A}_\pm^0 + \sum_{l=1}^L Z_\pm^l \log{(w - p_l)},~~~~Z^l_+=-\overline{Z^l_-}= \sigma \prod_{n=1}^{L-2}(p_l-s_n)\prod_{k\neq l}^L\frac{1}{p_l-p_k},
\end{equation}
where $\sigma$ is a complex constant. Here $p_l$ denote the poles of the differentials $\partial_w {\cal A}_\pm$, which lie on the real line. Note that here and elsewhere we define the logarithm such that
\beq
\log(-x)= i\pi+ \log(x).
\eeq
The condition ${\cal G}=0$ on the boundary holds if the parameters of the solution satisfy
\begin{equation}\label{eq:upperplnecond}
{\cal A}^0 Z^k_-+\bar{{\cal A}}^0 Z_+^k+\sum_{l\neq k}Z^{[l,k]}\log{|p_l-p_k|}=0,~~~~Z^{[l,k]}=Z^l_+Z^k_--Z^k_+Z^l_-
\end{equation}
for $k=1,\dots, L$, $2{\cal A}^0={\cal A}^0_+-\bar{{\cal A}^0_-}$, and one tunes an integration constant in ${\cal B}(w)$.
Interestingly, the behaviour of the supergravity fields close to the poles was shown to correspond to the near brane limit of $(p,q)$ 5-branes, 
with charges 
\begin{equation}
\label{q1q2}
(q_1^l-iq_2^l)=\frac43 Z_+^l\, ,
\end{equation}
in the notation in which $p\equiv q_2$ refers to the D5-brane charge\footnote{To avoid any confusion with the locations of the poles.}, and $q\equiv q_1$ to the NS5-brane charge. This relation identifies the charges of the 5-branes with the residues of $\partial_w {\cal A}_+$ at the poles $p_l$. 

\subsubsection{Addition of seven branes}\label{inclusion7branes}
The analysis in \cite{DHoker:2017zwj} revealed that the information about the branch cut structure of the punctures associated to 7-branes on the upper half-plane can be encoded in, 
\begin{equation}
\label{fomega}
f(w)=\sum_{i=1}^I \frac{n_i ^2}{4\pi}\log{\Bigl(\gamma_i \frac{w-w_i}{w-\bar{w}_i}\Bigr)},
\end{equation}
where $w_i$ is the locus of a 7-brane puncture, $\gamma_i$ is a phase specifying the orientation of the associated branch cut\footnote{Such that $\gamma_i$  is +1 for a branch cut extending in the negative imaginary direction and $-1$ for a branch cut in the positive imaginary direction.},
and $n_i^2$ is the number of 7-branes at that puncture - it is not hard to see that \eqref{fomega} indeed satisfies each condition in \eqref{eq:frules}.

Following the construction in \cite{DHoker:2017zwj} one arrives at the following expressions for 
${\cal A}_\pm$ for a solution with monodromy:
\begin{equation}
\label{Apm7branes}
{\cal A}_\pm = {\cal A}_\pm^0+\sum_{l=1}^L Y_\pm^l\log{(w-p_l)} +\eta_{\pm}\int_\infty^w dz\, f(z) \sum_{l=1}^L \frac{Y^l}{z-p_l},
\end{equation}
where
\beq \label{eq:Ydefs}
Y^l_{+}=u_{[p,q]}Z^l_+-v_{[p,q]}Z^l_-,~~~Y^l_-=-\overline{Y^l_+},~~~ Y^l= Z^l_+-Z^l_- \,,
\eeq
and the integration contour has been chosen such that it avoids crossing the poles on the real axis, as well as the punctures and associated branch cuts in $\Sigma$. The functions $u_{[r,s]}$ and $v_{[r,s]}$ are defined in eq.~\eqref{eq:monodromy} of section \ref{rs7branes}, where the action of the $SL(2,\mathbb{R})$ monodromies is introduced in detail.

In the presence of 7-branes, the vanishing of ${\cal G}$ on the entire real axis imposes that 
\beq\label{eq:upperplnecondd71}
0={\cal A}^0 {\cal Y}^k_-+\bar{ {\cal A}^0}{\cal Y}_+^k+\sum_{l\neq k}Y^{[l,k]}\log{|p_l-p_k|}+ \frac{1}{2}Y^k J_k,~~~k=1,...,L,
\eeq
where 
\begin{align}
J_k&= \sum_{l=1}^L Y^l\bigg[\int^{p_k}_{\infty}dx f'(x)\log|x-p_l|^2 +\sum_{ i \in {\cal S}_{[p_k,~\infty]}} \frac{i n_i^2}{2}\log|w_i-p_l|^2\bigg],\nn\\[2mm]
f'(x)&= \sum_{i=1}^I \frac{i n_i^2}{2 \pi}\frac{\text{Im}(w_i)}{|x-w_i|^2},~~~~ {\cal Y}^l_{\pm}= Y^l_{\pm}+ \eta_{\pm} f(p_l) Y^l,
\end{align}
and in $J_k$ the integration contour follows the real line. The final term takes account of deformations of this contour into the upper half-plane  to avoid crossing branch cuts that bisect $w\in\mathbb{R}$, as such ${\cal S}_{[p_k,~\infty]}$ is a partition of  $\{1,..I\}$ containing only the branch points whose cuts must be traversed in the interval $[p_l,~\infty)$. This ensures zero monodromy across the poles and generalises \eqref{eq:upperplnecond}. In addition to this, one must also impose the $SL(2,\mathbb{R})$ monodromy condition \eqref{eq: monodomycond}, which leads to 
\beq\label{eq:upperplnecondd72}
\eta_-{\cal A}^0 + \eta_+\overline{{\cal A}^0}+ \sum_{l=1}^L Y^l \log|w_i-p_l|^2=0,~~~~i=1,...,I.
\eeq
A global solution exists whenever one  solves \eqref{eq:upperplnecondd71} and \eqref{eq:upperplnecondd72}. Several examples can be found in \cite{DHoker:2017zwj,Gutperle:2018vdd}.


\subsection{The annulus}
\label{annulus}

In order to construct the differentials $\partial_w {\cal A}_\pm$ and the associated functions ${\cal A}_\pm$ for solutions with 5-branes on the annulus, DGU use the function theory on the double surface. 

The double surface ${\hat \Sigma}$ of the annulus is defined as a torus with periods 1 and $\tau$, with $\tau$ purely imaginary $\tau=it$, with $t\in \mathbb{R}^+$. ${\hat \Sigma}$ is then chosen by symmetry across the real axis, such that 
$0\leq {\rm Re}(w)\leq 1$ and $|{\rm Im}(w)|\leq t/2$. ${\hat \Sigma}$ is thus 
${\hat \Sigma}=\mathbb{C}/(\mathbb{Z}+\mathbb{Z}\tau)$.
The original surface $\Sigma$ is then obtained as the quotient $\Sigma=\hat{\Sigma}/{\cal J}$, where ${\cal J}$ maps the components of ${\hat \Sigma}$ in the upper and lower half-planes, and its boundary $\partial\Sigma$ is the fixed set under ${\cal J}$. 

The scalar Green function on the annulus is constructed from the scalar Green function on the double surface, and can be expressed in terms of a Jacobi theta-function of the first kind
\beq
\theta_1(w|\tau)= 2 \sum_{n=0}^{\infty} (-1)^n e^{i\pi(n+\frac{1}{2})^2\tau}\sin((2n+1)\pi w),
\eeq
as
\begin{equation}
G(w,s|\tau)=-\log{\left|\frac{\theta_1(w-s|\tau)}{\theta_1(w-{\bar s}|\tau)}\right|^2}+\frac{2\pi}{t}(w-{\bar w})(s-{\bar s}),
\end{equation}
where the well-known properties \cite{Green:2012pqa} of $\theta_1(w|\tau)$
\begin{eqnarray}
&&\theta_1(w+1|\tau)=-\theta_1(w|\tau)\nonumber\\ 
&&\theta_1(w+\tau|\tau)=-\theta_1(w|\tau)\exp{(-i\pi\tau-2\pi iw)},
\end{eqnarray}
ensure that $G$ vanishes at both boundary components of $\Sigma$, $\mathbb{R}$ and $\mathbb{R}+\frac{\tau}{2}$.

From $G$ the holomorphic function $\lambda(w|\tau)$ is then computed to be
\begin{equation}\label{eq:meromorphic_annulus}
\lambda(w|\tau)=\lambda_0^2\prod_{n=1}^N \frac{\theta_1(w-s_n|\tau)}{\theta_1(w-{\bar s}_n|\tau)}\times \exp{\Bigl\{-\frac{2}{t}w\sum_{n=1}^N (s_n-{\bar s}_n)\Bigr\}}
\end{equation}
where the points $s_n\in \Sigma$ are zeros of $\lambda$ and must satisfy $
\sum_{n=1}^N (s_n-{\bar s}_n)\in \mathbb{Z}\tau$, in order to achieve single-valuedness under $w\rightarrow w+1$.

Following in detail the derivation in \cite{DHoker:2017mds} the following expressions for ${\cal A}_\pm$ are obtained for the case in which all poles are on the real boundary component and there are no monodromies
\beq\label{eq:singlevaluedApmannulus}
{\cal A}_{\pm}={\cal A}^0_{\pm} + \sum_{l=1}^L Z^l_{\pm} \log{\theta_1(w-p_l|\tau)},~~~~Z^l_{+}=-\overline{Z^l}_{-}= \sigma \frac{\prod_{n=1}^L \theta_1 (p_l-s_n|\tau)}{\prod_{n\neq l}\theta_1(p_l-p_n|\tau)} \,\exp\left\{-\frac{2\pi i}{\tau}p_l\Lambda\right\},
\eeq
where $\sigma$ is a complex constant and
\beq\label{eq:condannulus1}
\Lambda=\sum_{n=1}^L(s_n-p_n) \in \mathbb{Z}\tau,
\eeq
which implies $
\sum_{n=1}^L (s_n-{\bar s}_n)\in \mathbb{Z}\tau$. 
The holomorphic functions \eqref{eq:singlevaluedApmannulus} are not actually a direct integration of the $\partial_{w}{\cal A}_{\pm}$, the recipe sketched in section \ref{sec:alaDHoker} leads to, rather they are complicated product expressions. In \cite{DHoker:2017mds} these are integrated after decomposing the products in a sum of Abelian differentials. This procedure generically leads to  ${\cal O}(w)$ terms in ${\cal A}_{\pm}$ which break the periodicity of ${\cal G}$ under $w\to w+1$, setting these terms to zero imposes
\beq\label{eq:condannulus2}
\sum_{l=1}^L Z^l_{+} \partial_s\log{\theta_1(s-p_l|\tau)}=0
\eeq
for $s$ an arbitrary individual zero $s_i$.

In \eqref{eq:singlevaluedApmannulus} the sum is over the number of poles of the solution, which are associated to $(q_1^l,q_2^l)$ 5-branes, with $Z^l_+$ defined as in equation (\ref{q1q2}). In turn, the integration constants ${\cal A}^0_\pm$  must satisfy
\begin{equation}\label{eq:condannulus3}
Z^k_- {\cal A}^0+Z^k_+ \overline{{\cal A}^0}+\sum_{l\neq k}Z^{[l,k]}\log{|\theta_1(p_k-p_l)|}=0,
\end{equation}
where $2{\cal A}^0= {\cal A}^0_+-\overline{{\cal A}^0_-}$, to ensure that there is no monodromy around the poles. As with the upper half-plane, ${\cal G}=0$ for $w\in\mathbb{R}$ can be achieved by tuning the integration constant in ${\cal B}(w)$, but one also needs to impose that  ${\cal G}=0$ for $w\in\mathbb{R}+\frac{\tau}{2}$. This leads to the additional constraint
\begin{align}\label{eq:zeroGtauboundary}
\frac{{\cal G}(x+\frac{\tau}{2})}{2\pi i}&=2\sum_{l=1}^L({\cal A}^0 Z^l_-+\overline{{\cal A}^0}Z^l_+)p_l+\sum_{l\neq l'}^LZ^{[l,l']}\bigg[-p_l \log\theta_1(p_{l'}+\frac{\tau}{2}|\tau)\nn\\[2mm]
&+\frac{1}{2\pi i}\bigg(\int_0^{\frac{\tau}{2}}dw \log\theta_1(w-p_l|\tau)\partial_w \log\theta_1(w-p_{l'}|\tau)-\text{c.c}\bigg)\bigg]=0.
\end{align}
The details of this computation can be found in Appendix \ref{Annulusappendix}. So far no globally regular solutions on the annulus have been found. In reference \cite{DHoker:2017mds} a numerical study was performed which confirmed that for $L\geq 3$ it is possible to solve \eqref{eq:condannulus1}-\eqref{eq:condannulus3}, but in all attempted cases \eqref{eq:zeroGtauboundary} was not satisfied.
\vspace{0.3cm}


\section{Realising  7-branes in the annulus}
\label{rs7branesannulus}
In this section we consider adding 7-branes to the annulus set-up of the previous section. Our analysis  is actually not fully general, as we are assuming that there are no poles located in $\mathbb{R}+\frac{\tau}{2}$ and that a potential $ \mathcal{O}(w)$ term in ${\cal A}_{\pm}$ vanishes. We relax these assumptions in  Appendix \ref{eq:annulusD7appendix} where many additional technical details can be found.

A function with the required properties to add 7-branes to the annulus is
\beq
\label{f-annulus}
f(w) = \sum_{i=1}^I\frac{ n_i^2}{4 \pi}\bigg(\log\left(\gamma_i \frac{\theta_1(w-w_i|\tau)}{\theta_1(w-\overline{w_i}|\tau)}\right)- \frac{2\pi i }{\tau} (w_i-\overline{w_i})w\bigg),
\eeq
which is to say that in addition to satisfying \eqref{eq:frules}, it is also such that $f(w) + \overline{f(w)}$ is periodic under $w\to w+1$, from which it follows that $\kappa^2$  will also have this property. 

In order to have a well-defined solution on the annulus with localised 5 and 7-branes, there are  several conditions that need to be satisfied. First of all, two conditions are inherited from the seed solution - the annulus without 7-branes:
\beq\label{eq:seedcond}
\sum_{n=1}^L (s_n-p_n)\in \mathbb{Z}\tau,~~~~\sum_{l=1}^L Z^l_{+} \partial_s\log{\theta_1(s-p_l|\tau)}=0,
\eeq
for $s\in\{s_1,...s_L\}$, which essentially ensure that holomorphic functions of the seed solution are periodic under $w\to w+1$. The holomorphic functions with 7-branes will no longer respect this periodicity but this is not necessarily an issue\footnote{It is sufficient that the physical fields (metric, dilaton and gauge invariant fluxes) have this property.}. As discussed  at greater length in Appendix  \ref{eq:annulusD7appendix}, accommodating the $w\to w+1$ period requires us to restrict our considerations to the addition of (1,0) 7-branes, ie standard D7-branes. Thus the holomorphic functions we consider are 
\beq\label{eq:annulusD7}
{\cal A}_{\pm} ={\cal  A}_{\pm}^0 +\sum_{l=1}^L Z^l_{\pm} \log(\theta_1(w-p_l|\tau))+\int_1^w dz f(z)\sum_{l=1}^L Y^l\partial_{z}\log(\theta_1(z-p_l|\tau))
\eeq
with  $Y^l=Z^l_+-Z^l_-$ for the $Z^l_{\pm}$ of section \ref{annulus},  $w=1$ is a reference point that is assumed to be regular, and we fix
\beq\label{eq:acondand7}
{\cal A}^0_{\pm}=-\overline{{\cal A}^0_{\mp}}.
\eeq
Restricting to D7's is not actually sufficient to have periodicity under $w\to w+1$. For that we must impose that ${\cal A}_{\pm}(w+1)$ is related to  ${\cal A}_{\pm}(w)$ by an $SL(2,\mathbb{R})$ transformation of the form \eqref{eq:monodromy}, so that the physical fields at $w$ equal those at $w+1$. This requires fixing
\beq
\text{Im} \int^1_0 dw {\cal A}^{(0)}(w)\partial_w f(w)=0,
\eeq 
which ensures that the constant components of the transformations of ${\cal A}_{\pm}(w)$ do indeed come in a complex conjugate pair as in \eqref{eq:monodromy}.
We can simplify this expression by localising the integration contour to the real axis, with deformations to avoid branch cuts that cross this path - these deformations only contribute at the poles of $\partial_w f(w)$ - they make up the second term in ${\cal K}$ defined below. We then find that one real constraint
\begin{equation}\label{eq:periodcond}
0=\frac{2 i }{\tau}({\cal A}^0_+-{\cal A}^0_-) \sum_{i=1}^I(n_i)^2\text{Im}(w_i)+{\cal K},
\end{equation}
with
\begin{equation}
{\cal K}=\sum_{l=1}^LY^l\bigg[i\int^0_1 dx \log|\theta_1(x-p_l|\tau)|^2f'(x)-\frac{1}{2}\sum_{i\in {\cal S}_{[0,1]}}(n_i)^2\log|\theta_1(w_i-p_l|\tau)|^2\bigg],
\end{equation}
is required to ensure $w\to w+1$ periodicity, where the integral is now performed over the whole real axis without deformation.

Ensuring that a suitable $SL(2,\mathbb{R})$ monodromy is generated as we move around each branch point requires that \eqref{eq: monodomycond} is satisfied. This gives $I$ additional constraints
\beq\label{eq:su2monocond}
2({\cal A}^0_+-{\cal A }^0_-)+ \sum_{l=1}^L Y^l \log|\theta_1(w_i-p_l|\tau)|^2=0,
\eeq
one for each branch point. In turn, ensuring that there is zero monodromy across the poles on the real axis, gives $L-1$ additional independent constraints of the form
\begin{equation}\label{eq:monocond}
0=2({\cal Y}^k_- {\cal A}^0_+-{\cal Y}^k_+{\cal A}^0_-) +\sum_{l\neq k}Z^{[l,k]}\log{|\theta_1(p_k-p_l)|^2}+Y^k\bigg(2\sum_{l=1}^LZ^l\theta_1(p_l|\tau)f(1)+J_k)\bigg),
\end{equation}
with\footnote{We remind the reader that  ${\cal S}_{[p_k,1]}$ contains only the $i\in\{1,...,I\}$ that correspond to  branch cuts that are crossed following a straight path between $p_k$ and $1$.}
\begin{equation}
J_k= \sum_{1=1}^LY^l\bigg[\int^{p_k}_1dx\log|\theta_1(x-p_l|\tau)|^2 f'(x)+\frac{i}{2}\sum_{i\in {\cal S}_{[p_k,1]}}(n_i)^2\log|\theta_1(w_i-p_l|\tau)|^2\bigg],
\end{equation}
and
\beq
{\cal Y}^k_{\pm}= Z^l_{\pm}+ Y^l f(p_k).
\eeq 
The additional term, with respect to the upper half-plane with 7-branes, that is proportional to $f(1)$, is actually quite generic, its coefficient was just set to zero with a clever choice of reference point on the plane.

The final thing we need to ensure is that ${\cal G}$ vanishes on the two boundaries. For $w\in \mathbb{R}$, as shown in Appendix \ref{eq:annulusD7appendix}, this amounts, given \eqref{eq:acondand7}, to tuning the integration constant in ${\cal B}(w)$ as
\beq
\text{Re }{\cal B}^0=0,
\eeq 
which  fixes ${\cal G}(w)=0$ on the real axis. For $w\in \mathbb{R}+\frac{\tau}{2}$  things are once more a little more involved. Following the derivation in Appendix \ref{Annulusappendix} to get to \eqref{eq:genannulusupperbound} and integrating by parts, we find that the vanishing of ${\cal G}(w)$ on the second boundary requires
\begin{align}\label{eq:zeroGtauboundaryD7}
0&=2\sum_{l=1}^L({\cal A}^0_+ X^l_--{\cal A}^0_-X^l_+)p_l+\sum_{l\neq l'}^LZ^{[l,l']}\bigg[-p_l \log\theta_1(p_{l'}+\frac{\tau}{2}|\tau)\nn\\[2mm]
&+\frac{1}{\pi }\text{Im}\bigg(\int_0^{\frac{\tau}{2}}dw \log\theta_1(w-p_l|\tau)\partial_w \log\theta_1(w-p_{l'}|\tau)\bigg)\bigg]\nn\\[2mm]
&+\sum_{l,l'=1}^LY^l Y^{l'}\bigg[2p_lf(1)\log\theta_1(p_{l'}|\tau)+i p_l\text{Im}\bigg(\int_0^{\frac{\tau}{2}}dw\log\theta_1(w-p_{l'}|\tau)\partial_{w}f(w)\bigg)\nn\\[2mm]
&+\frac{i}{\pi}\text{Re}\bigg(\int_0^{\frac{\tau}{2}}dw \log\theta_1(w-p_l)\log\theta_1(w-p_{l'})\partial_w f(w)\bigg)\nn\\[2mm]
&+\frac{i}{2}\sum_{i\in S_{[0,\frac{\tau}{2}]}}s_i(n_i)^2 \bigg(p_l\text{Re}\log\theta_1(w_i-p_{l'}|\tau)-\frac{1}{\pi}\text{Im}\log\theta_1(w_i-p_{l}|\tau)\log\theta_1(w_i-p_{l'}|\tau)\bigg)\bigg]
\end{align}
where we define
\beq
X_{\pm}^l=Z^l_{\pm}+Y^l f(1+\frac{\tau}{2})+\frac{i Y^l}{\pi} \text{Re}\int_0^{\frac{\tau}{2}}dw \log\theta_1(w-p_l)\partial_w f(w)-\frac{iY^l}{2\pi} \sum_{i\in S_{[0,\frac{\tau}{2}}]}s_i(n_i)^2\text{Im}\log\theta_1(w_i-p_l)\nn.
\eeq
As before we have split the full contour into a straight line contribution from $0$ to $\frac{\tau}{2}$, and a contribution from the poles in $\partial_w f$.  As such
\beq
s_i=\pm 1
\eeq
with the $+$  taken when a pole is traversed anti-clockwise, and $-$ taken for the converse.

In conclusion, if one is able to simultaneously solve equations \eqref{eq:seedcond}, \eqref{eq:periodcond}, \eqref{eq:su2monocond}, \eqref{eq:monocond} and \eqref{eq:zeroGtauboundaryD7}, one will have a compact solution on the annulus with localised 5 and 7-branes. Although this appears to be an intimidating task, we note that at least in the case of the upper half-plane the additional monodromy conditions were not difficult to solve. Additionally, the barrier to constructing compact solutions with localised 5-branes on the annulus was getting ${\cal G}$ to vanish on both boundaries. We hope that the additional 7-brane terms appearing in equation \eqref{eq:zeroGtauboundaryD7} can resolve this issue. We leave a detailed study of this to future work.

\section{The $AdS_6$ T-duals of Brandhuber-Oz}
\label{sec:gravity_solutions}

The first $AdS_6$ solutions to Type IIB supergravity were constructed in 
\cite{Lozano:2012au,Lozano:2013oma} by acting with Abelian and non-Abelian T-duality on the Brandhuber-Oz solution to massive IIA. In the next sections we will show that these solutions fit within an extension of the global $AdS_6$ solutions discussed in the previous sections. 
Prior to that, we review in this section the main properties of the T-dual solutions, starting with a brief summary of  the BO solution.

\subsection{The Brandhuber-Oz $AdS_6$ solution}
\label{sec:BO}

The Brandhuber-Oz solution \cite{Brandhuber:1999np} describes the near-brane region of the Type IIA  configuration consisting on $N$ D4-branes probing a $O8^-$ orientifold fixed plane with $N_f$ coincident D8-branes. This configuration realises in string theory  the strongly coupled UV fixed point associated to the 5d  $USp(2N)$ gauge theory with one antisymmetric hypermultiplet and $N_f<8$ fundamental hypermutiplets, considered in \cite{Seiberg:1996bd,Intriligator:1997pq,Ferrara:1998gv}. The underlying brane set-up is summarised in Table \ref{branesBO}.
\begin{table}[ht]
	\begin{center}
		\begin{tabular}{| l | c | c | c | c| c | c| c | c| c | c |}
			\hline		    
			& 0 & 1 & 2 & 3 & 4 & 5 & 6 & 7 & 8 & 9 \\ \hline
			D4 & x & x & x & x & x &  &   &   &   &   \\ \hline
			D8/O8 & x & x & x & x & x & x  & x & x & x &   \\ \hline
		\end{tabular} 
	\end{center}
	\caption{Brane intersection associated to the Brandhuber-Oz solution.}   
	\label{branesBO}	
\end{table} 

The worldvolume theory of the D4-branes (plus their images) is the $USp(2N)$ gauge theory. The scalar in the vector multiplet corresponds to the positions of the D4-branes in $x^9$, and parameterises the Coulomb branch of the theory. The antisymmetric hypermultiplet corresponds to the positions of the D4-branes in $(x^5,x^6,x^7,x^8)$ and parameterises, together with the $N_f$ fundamental hypermultiplets coming from the D4-D8 strings, the Higgs branch. There is a global $SU(2)\times SO(2N_f)$ symmetry associated to the antisymmetric hypermultiplet and the coincident $N_f$ D8-branes, together with a $U(1)$ instantonic symmetry associated to the 10d RR 1-form potential. The second $SU(2)$ in the $SO(4)$ rotation symmetry in $(x^5,x^6,x^7,x^8)$ corresponds to the $SU(2)$ R-symmetry of the 5d $\mathcal{N}=1$ gauge theory. The theory at the origin of the Coulomb branch is a 5d fixed point, where the global symmetry is enhanced to $SU(2)\times E_{N_f+1}$ \cite{Seiberg:1996bd}.

The Brandhuber-Oz solution is a warped product of $AdS_6$ times a half-$S^4$, that can be expressed as
\vspace{-2mm}
\begin{align}\label{eq:AdS6 sol}
ds^2&= \frac{L^2 W^2}{4}\bigg[9 ds^2(AdS_6)+ 4\bigg(d\theta^2+\sin^2\theta ds^2(S^3)\bigg)\bigg],\qquad e^{\Phi}=\frac{2}{3L} W^{5},\nn\\[2mm]
F_0&=m,\quad F_4 = \frac{5L^4}{W^2}\sin^3\theta d\theta\wedge \text{Vol}(S^3),\quad W= (m \cos\theta)^{-\frac{1}{6}},
\end{align}
where $L$ denotes the $AdS_6$ radius and $\theta\in [0,\frac{\pi}{2}]$. In the near-brane limit the $\theta$-direction and the radius of $AdS_6$ emerge as functions of the $x^5,\dots,x^9$ directions transverse to the D4-branes. The orientifold action, $x^9\rightarrow -x^9$, becomes
$\theta\rightarrow \pi-\theta$, and the orientifold fixed plane is located at $\theta=\pi/2$, where the solution is singular. This singularity is consistent with a D8/O8 system, since defining $v^3=(\pi/2-\theta)^{2}$, we have, near $\theta=\frac{\pi}{2}$:
\begin{equation}
ds^2 \sim \frac{L^2}{4m^{1/3}} \bigg[ \frac{1}{\sqrt{v}}\bigg( 9ds^2(AdS_6) + 4ds^2(S^3)\bigg) + \sqrt{v}\, 9dv^2 \bigg] \,, \quad e^\Phi\sim\frac{2}{3Lm^{5/6}v^{5/4}}\;.
\end{equation}
The solution covers a half-$S^4$, and therefore only one side of the orientifold fixed plane.  
It exhibits the explicit $SO(2,5)\times SU(2)_R$ bosonic symmetry of $F(4)$, plus the global $SU(2)$ symmetry associated to the antisymmetric hypermultiplet. The flavour symmetry however is not shown, which is consistent with the fact that this symmetry occurs at the origin of the Coulomb branch, where the solution is singular \cite{Bergman:2012kr}.

Flux quantisation requires that
\beq\label{eq:BO_charges}
m = \frac{1}{2\pi}N_{8},~~~ L^4=\frac{16\pi}{9}\left(\frac{2\pi}{N_{8}}\right)^{1/3}\!\!\!\! N_{D4},
\eeq
where $m$ is the Romans mass and $N_{8}\equiv 8-N_f$ accounts for the charge of the O8 plane and the $N_f$ D8-branes. $N_8$ and $N_{D4}$ must be integers and $N_f \leq 7$. We use the conventions
\beq
N\equiv N_{D4}=\frac{1}{2\kappa_{10}^2T_4}\int_{S^4}F_4,~~~N_{8}=\frac{m}{2\kappa_{10}^2T_8},
\eeq
and $ 2\kappa_{10}^2T_p= (2\pi)^{7-p}$. 

The supergravity limit is valid when
\beq
N_{8}^3 N_{D4}>>1,~~~~ \frac{N_{D4}}{N_{8}}>>1\,.
\eeq

\subsection{The $AdS_6$ Abelian T-dual  (ATD) solution}
\label{ATDsub}

The $USp(2N)$ theory with one antisymmetric hypermultiplet and $N_f<8$ fundamental hypermultiplets can also be realised in Type IIB string theory. In Type IIB the intersection consists of  two O7 planes spanned on $\mathbb{R}^{1,4}\times \mathbb{R}^3$, located at $x^5=0,\pi/2$, $N/2$ D5-branes spanned on $\mathbb{R}^{1,4}\times S^1$, and $N_f/2$ D7-branes parallel to the O7-planes, plus their mirrors. The brane configuration is summarised in Table \ref{branesATD}.
\begin{table}[ht]
	\begin{center}
		\begin{tabular}{| l | c | c | c | c| c | c| c | c| c | c |}
			\hline		    
			& 0 & 1 & 2 & 3 & 4 & 5 & 6 & 7 & 8 & 9 \\ \hline
			D5 & x & x & x & x & x & x &   &   &   &   \\ \hline
			D7/O7 & x & x & x & x & x &   & x & x & x &   \\ \hline
			NS5 & x & x & x & x & x &   &   &   &  & x \\ \hline
		\end{tabular} 
	\end{center}
	\caption{Brane intersection associated to the Abelian T-dual of BO. In the near brane limit $x^6$ and $x^9$ mix \cite{Bergman:2012kr}, to produce D7/O7 branes spanned on $AdS_6\times S^2$ and NS5-branes spanned on $AdS_6$, consistently with the isometries of the background. These branes are located at $\theta=\pi/2$ and $\theta=0$, respectively.}   
	\label{branesATD}	
\end{table} 

In the near brane limit the previous brane intersection gives rise to the T-dual of the Brandhuber-Oz solution \cite{Lozano:2012au,Lozano:2013oma},
\begin{align}\label{eq:AdS6_sol_ATD}
ds^2&= \frac{L^2 W^2}{4}\bigg[9 ds^2(AdS_6)+ \sin^2\theta ds^2(S^2)\bigg]+\frac{4}{L^2W^2\sin^2\theta}\left(d\psi^2+\frac{1}{4}L^4W^4\sin^2\theta d\theta^2\right) ,\nn\\[2mm]
e^{-\Phi}&=\frac{3L^2}{4W^4}\sin\theta,~~~B_2=\cos{\xi}d\psi\wedge d\phi,  
~~~F_1=-m d\psi,\\[2mm] 
F_3&=\frac{5 L^4}{8 W^2}\sin^3\theta d\theta\wedge\text{Vol}(S^2),~~~~~ F_5=0.\nn
\end{align}
In this limit $x^5$ becomes the $\psi$ direction, $\psi \in [0,\pi]$, and the $x^6$ and $x^9$ directions combine to produce the $AdS_6$ radius and the $\theta$ direction. In turn, $(\xi,\phi)$ parameterise the $S^2\subset \mathbb{R}^3$ associated to the $SU(2)$ R-symmetry of $F(4)$, such that $\text{Vol}(S^2)=\sin{\xi}d\xi \wedge d\phi$. In the T-duality transformation the global $SU(2)$ symmetry associated to the antisymmetric hypermultiplet is reduced to $U(1)$. This symmetry is however enhanced to $SU(2)$, in agreement with the enhanced mesonic symmetry (see \cite{Bergman:2012kr}).

The solution described by  (\ref{eq:AdS6_sol_ATD}) has a singularity at $\theta=\frac{\pi}{2}$, inherited from the D8/O8 system in IIA. Close to this point
\beq
ds^2 \sim \frac{L^2}{4m^{1/3}} \bigg[ \frac{1}{\sqrt{v}}\bigg(9 ds^2(AdS_6)+ ds^2(S^2)\bigg)  +  9\sqrt{v}\bigg(dv^2 + d\tilde{\psi}^2\bigg) \bigg]\,, \quad e^\Phi\sim \frac{4}{3L^2m^{2/3}v} \;,
\eeq
where $v^3=(\frac{\pi}{2}-\theta)^2$ and $3L^2\tilde \psi= 4m^{1/3}\psi$. This is the behaviour of the smeared D7/O7 system that arises after T-duality. Further, in this case there is a second singularity at $\theta=0$ where we find
\beq
ds^2 \sim \frac{L^2}{4m^{1/3}} \bigg[ 9 ds^2(AdS_6)  +  \frac{1}{v}\bigg(dv^2 + d\tilde{\psi}^2+v^2ds^2(S^2) \bigg) \bigg]\,, \quad e^\Phi\sim \frac{4}{3L^2m^{2/3}\sqrt{v}} \;,
\eeq
where $v^2= \theta$ and $L^2\tilde \psi= 4m^{1/3}\psi$. This is the behaviour of the smeared NS5 branes one generates when T-dualising on the Hopf fibre of a 3-sphere that shrinks at a certain point.

Flux quantisation requires that
\beq\label{eq:ATD_charges}
m = \frac{1}{\pi}N_{7},~~~ L^4=\frac{32\pi}{9}\left(\frac{\pi}{N_{7}}\right)^{1/3} N_{D5}
\eeq
so that the Page charges
\begin{align}\label{eq:Page_charges_ATD}
N_{7}&=\frac{1}{2\kappa_{10}^2 T_7}\int_{\psi=0}^{\psi=\pi} F_1=-m\pi ,\nn\\[2mm]
N_{D5}&=\frac{1}{2\kappa_{10}^2 T_5}\int_{S^2}\int_{\theta=0}^{\theta=\frac{\pi}{2}} (F_3-B_2\wedge F_1)=\frac{N_{D4}}{2},\nn\\[2mm]
N_{NS5}&=\frac{1}{2\kappa_{10}^2 T_5}\int_{S^2}\int_{\psi=0}^{\psi=\pi} H_3 =1
\end{align}
are all integers. 
Note that the T-duality transformation maps the O8 plane onto two O7 planes with charge 4, and the $N_f$ D8-branes onto $N_f/2$ D7-branes (plus their mirrors). $N_7$ accounts for the charge of each D7/O7 system. NS5-branes are created as well at the point where the original $S^1$ shrinks.

The supergravity limit is valid when 
\beq
N_{7} N_{D5}>>1,~~~\frac{N_{D5}}{N_{7}}>>1.
\eeq


\subsection{The $AdS_6$ non-Abelian T-dual (NATD) solution}
\label{NATD}

The Abelian T-dual solution provides the holographic dual in Type IIB string theory of the same UV fixed point theory associated to the BO solution. This is guaranteed by the fact that Abelian T-duality is a string theory symmetry. The usage of non-Abelian T-duality as a solution generating technique allows, in principle, for more interesting possibilities, given that in contrast to its Abelian counterpart, there is no proof that it stands as a string theory symmetry.  Previous examples in a holographic context \cite{Lozano:2016kum,Lozano:2016wrs,Lozano:2017ole,Itsios:2017cew} indeed suggest that non-Abelian T-duality may change the CFT dual of the AdS solutions on which it is applied.

Non-Abelian T-duality was first used in the context of 5d fixed point theories in \cite{Lozano:2012au}. In this reference the first  explicit $AdS_6$ solution to Type IIB supergravity\footnote{Other than the Abelian T-dual of BO, just discussed.} was constructed, showing that this type of solutions could exist, contrary to the expectations from \cite{Passias:2012vp}.
This prompted the investigation of general classifications of $AdS_6$ solutions to Type IIB supergravity, that culminated with the work of \cite{DHoker:2017mds,DHoker:2017zwj,D'Hoker:2016rdq,DHoker:2016ysh}, summarised in section 2.

The explicit solution constructed in \cite{Lozano:2012au} was obtained dualising the BO solution with respect to one of the $SU(2)$ isometry groups of the  
half-$S^4$. 
Using the conventions in  \cite{Lozano:2013oma} it reads
\begin{align}\label{eq:AdS6_sol_NATD}
ds^2&= \frac{L^2 W^2}{4}\bigg[9 ds^2(AdS_6)+ \frac{r^2}{\Delta}\sin^2\theta ds^2(S^2)\bigg]+\frac{4}{L^2W^2\sin^2\theta}\left(dr^2+\frac{1}{4}L^4W^4\sin^2\theta d\theta^2\right) ,\nn\\[2mm]
e^{-\Phi}&=\frac{3L^2}{4W^4}\sin\theta\sqrt{\Delta},~~~B_2=\frac{r^3}{\Delta}\text{Vol}(S^2),~~~F_1=\frac{5L^4}{8 W^2}\sin^3\theta d\theta- m r dr ,\nn\\[2mm]
F_3&=\frac{L^4 r^2\sin^3\theta}{16 \cos\theta W^2 \Delta}\big(-10r \cos\theta d\theta+\sin\theta dr\big)\wedge \text{Vol}(S^2),~~~F_5=0,~~~\Delta=r^2+ \left(\frac{LW\sin\theta }{2}\right)^4.
\end{align}
In this solution $r\in \mathbb{R}^+$, and is thus non-compact. This is common to all non-Abelian T-dual solutions constructed with respect to a freely acting $SU(2)$ isometry group, and has to do with the fact that under this operation the $SU(2)$ isometry group is replaced by its $\mathbb{R}^3$ Lie algebra. The solutions thus need to be completed globally such that they can describe holographically well-defined CFTs. This will be the subject of our discussion in section \ref{field-theory}.

As the Abelian T-dual, the non-Abelian T-dual solution
has two singularities. The  first one is inherited from the BO solution, and is thus associated to the location of the non-Abelian T-dual of the O8 fixed plane. Expanding around $\theta= \pi/2$ we find that
\begin{equation}\label{eq:NATD_near_D5}
ds^2 \sim \frac{9L^2}{4m^{1/3}} \bigg[ \frac{1}{\sqrt{v}} ds^2(AdS_6)  +  \sqrt{v}\bigg(dv^2 + d\tilde{r}^2+\tilde{r}^2ds^2(S^2)\bigg) \bigg]\,, \quad e^\Phi\sim \frac{16}{3L^4m^{1/3}\sqrt{v}} \;,
\end{equation}
where  $v^3=(\pi/2-\theta)^{2}$ and $3L^2\tilde r= 4m^{1/3}r$. This is the behaviour of D5-branes smeared on $\mathbb{R}^3$, which suggests that under non-Abelian T-duality the D8/O8 system is mapped onto a D5/O5 system, smeared in these directions. This is in agreement with the worldsheet analysis performed in Appendix \ref{worldsheet}, which shows that worldsheet parity reversal is mapped under non-Abelian T-duality (under a freely acting $SU(2)$) onto the combined action of this operation and an inversion in $\mathbb{R}^3$. The O8 fixed plane is thus mapped onto an O5 fixed plane, located at $r=0$.
As the BO solution, the non-Abelian T-dual solution just describes the 
$\theta\in [0,\pi/2]$, $r\in \mathbb{R}^+$ physical region. We will see in what follows that it will be more convenient to think of the O5 fixed plane as a O7 fixed plane wrapped on a collapsing  $S^2$ at $r=0$. 

The second singularity present in the non-Abelian T-dual solution is at $\theta=0$, where we find
\begin{equation}\label{eq:NATD_near_NS5}
ds^2 \sim \frac{L^2}{4m^{1/3}} \bigg[ 9 ds^2(AdS_6)  +  \frac{1}{v}\bigg(dv^2 + d\tilde{r}^2+v^2ds^2(S^2) \bigg) \bigg]\,, \quad e^\Phi\sim \frac{16}{3L^4m^{1/3}\tilde{r}\sqrt{v}} \;,
\end{equation}
for $v= \theta^2$ and $L^2\tilde r= 4m^{1/3}r$. This is consistent with NS5-branes smeared on ${\tilde r}$. This is very similar to what we found for the Abelian T-dual solution, since also in this case the $S^3$ along which we dualise shrinks to a point at this value of $\theta$. Note that the $SU(2)$ global symmetry associated to the antisymmetric hypermultiplet of the $USp(2N)$ theory has disappeared in the new solution. This has implications that we will discuss in section \ref{field-theory}.

\vspace{0.5cm}

Let us now analyse the quantised charges associated to the solution. Looking at the $B_2$ field, we see that at $\theta=0$, $B_2= r \text{Vol}(S^2)$. Therefore, as we move in $r$ we should impose that  \cite{Lozano:2013oma} 
\beq
b_0=\frac{1}{4\pi^2} \int_{S^2}B_2,~~~ 0\leq b_0<1.
\eeq
This is achieved through a large gauge transformation of parameter $n$: $B_2\rightarrow B_2 - n\pi \text{Vol}(S^2)$, for
$n\pi<r<(n+1)\pi$. The non-compactness of $r$ is thus reflected in the existence of large gauge transformations of infinite gauge parameter $n$.
Taking into account these large gauge transformations, the quantised charges in the interval $n\pi<r<(n+1)\pi$ become
\begin{align}\label{eq:Page_charges_NATD}
N_{D7}&=\frac{1}{2\kappa_{10}^2 T_7}\int_{\theta=0}^{\theta=\frac{\pi}{2}} F_1=\frac{9}{32} L^4 m^{1/3} ,\\[2mm]
\label{D5charge}
N_{D5}&=\frac{1}{2\kappa_{10}^2 T_5}\int_{S^2}\int_{\theta=0}^{\theta=\frac{\pi}{2}} (F_3-B_2\wedge F_1)=n N_{D7}, \\[2mm]
\label{ND7r} 
N^r_{7}&=\frac{1}{2\kappa_{10}^2 T_7}\int_{r=n\pi}^{r=(n+1)\pi} F_1=-m\pi^2 (n+\frac12),\\[2mm]
\label{ND5r}
N^r_{5}&=\frac{1}{2\kappa_{10}^2 T_5}\int_{S^2}\int_{r=n\pi}^{r=(n+1)\pi} (F_3-B_2\wedge F_1) = m\pi^2 (\frac{n}{2}+\frac13)  , \\[2mm]
N^n_{NS5}&=\frac{1}{2\kappa_{10}^2 T_5}\int_{S^2}\int_{r=n\pi}^{r=(n+1)\pi} H_3 =1 .
\label{NScharge}
\end{align}
In this case we have the condition
\beq\label{eq:NATD_charges}
L^4= \frac{32}{9}\frac{N_{D7}}{m^{1/3}}\, . 
\eeq


The existence of five different quantised charges seems to imply that extra branes have been created in the non-Abelian T-dual solution. However, as in the Abelian T-dual one, the number of independent charges remains equal to 3. The difference with respect to the Abelian case is that NS5-branes keep being created as we move in $r$, according to equation (\ref{NScharge}), due to its non-compactness.
In a given interval $n\pi<r<(n+1)\pi$ the $n$ $B_2$ charge has the effect of dissolving D5-brane charge on the D7-branes, thus duplicating the RR quantised charges associated to the solution.

A detailed relation between non-Abelian T-dual solutions constructed by dualising freely acting $SU(2)$ isometry groups and Abelian T-duals constructed with respect to their $U(1)$ fibres, was pointed out in \cite{Lozano:2016kum} (see also \cite{Lozano:2016wrs}). In these references it was noticed that the non-Abelian T-dual solution reduces to the Abelian T-dual one if one compactifies $r$ in $[n\pi,(n+1)\pi]$ and takes the large $n$ limit. In this limit $r$ plays the role of the  $\psi\in [0,\pi]$ $U(1)$ direction  of the Abelian T-dual solution, and both solutions agree up to an $r^2$ factor in the dilaton that has to do with their two different integration measures (see \cite{Lozano:2016kum} for more details).  In this limit the charges of the D-branes present in both solutions are related through $N^{A}_i=n\pi N^{NA}_i$.  

The relation with the Abelian T-dual solution suggests that in the non-Abelian case the brane intersection that emerges in the near brane limit could originate from the same type of branes involved in the supersymmetric brane system underlying the Abelian T-dual solution, shown in Table \ref{branesATD}. In the non-Abelian case, there are redundant charges as compared to the Abelian T-dual, that could have their origin in a more complicated mixing between the $(x^6,x^9)$ directions after the near horizon limit is taken.
In the large $r$ limit, when $r=\psi$, $r$ would  be identified with the  {\it field theory direction} of the dual field theory, as in the Abelian T-dual case.  In this limit the $(x^6,x^9)$ directions would simply mix to produce NS5-branes spanned on $AdS_6$ and D7/O7 branes spanned on $AdS_6\times S^2$, as in that case. 

Given our lack of knowledge about how D-branes transform under non-Abelian T-duality\footnote{See  \cite{Borlaf:1996na} for some efforts in this direction and \cite{Driezen:2018glg} for more recent ones.}, an underlying (D5, NS5, D7/O7) brane system
is only known to be  correct in the large $r$ limit. Still, beyond that limit it is possible to use some arguments that support it based on previous analysis of non-Abelian T-dual solutions. 

Indeed,  
in previous examples of non-Abelian T-duals \cite{Lozano:2016kum,Lozano:2016wrs,Itsios:2017cew} it was possible to use additional information coming from their specific geometries that supported the brane configurations inferred directly from the solutions. The best stablished cases are the $AdS_5\times S^2$ and $AdS_4\times S^2\times S^2$ examples studied in \cite{Lozano:2016kum} and \cite{Lozano:2016wrs}, respectively. The realisation of the first case as a Gaiotto-Maldacena geometry \cite{Gaiotto:2009gz} and of the second one as a Assel-Bachas-Estes-Gomis geometry \cite{Assel:2011xz} confirmed that the respective brane set-ups consisted on the same type of branes underlying the solutions arising in the corresponding Abelian T-dual limits. Based on these observations we think it is reasonable to expect a similar scenario  in this case. This will be the basis of our analysis in section \ref{sec:NATD_DGU} when we reproduce the non-Abelian T-dual solution as a DGU geometry.

\section{The Abelian T-dual as a DGU solution for the annulus}
\label{Abelian-T-dual}

In this section we show that the Abelian T-dual of the Brandhuber-Oz solution
fits within our extension in section 3 of the global supersymmetric $AdS_6$ solutions to type IIB supergravity found in \cite{DHoker:2017mds,DHoker:2017zwj}. The construction involves a Riemann surface $\Sigma$ with two boundaries and the topology of an annulus, which contains as well monodromies associated to 7-branes. This provides the first explicit example for the annulus, with the crucial addition of smeared D7/O7 branes. Locally, it was shown in \cite{D'Hoker:2016rdq} that the Abelian T-dual solution solves the set of equations derived therein, thus providing a useful consistency check of these equations\footnote{Even if being a singular solution it breaks some of their regularity conditions.}. 

\begin{figure}
	\centering
	\includegraphics[scale=1.3]{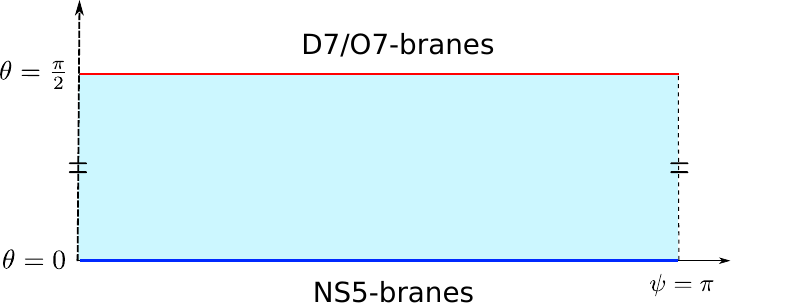}
	\caption{The annulus for the Abelian T-dual background. NS5-branes are smeared along the lower boundary at $\theta=0$, and 
		D7/O7 branes are smeared at the upper boundary at $\theta=\pi/2$. The annulus topology follows from the periodicity under $\psi\rightarrow \psi+\pi$, or $w\rightarrow w+1$, shifts.}
	\label{fig:annulus_ATD}
\end{figure}

The transverse $(\psi,\theta)$-directions of the Abelian T-dual solution parameterise a two dimensional Riemann surface $\Sigma$ which, given the periodicity of the $\psi$ direction, has the topology of an annulus. Taking the local holomorphic coordinate $w$ to be\footnote{With this choice the periodicity under $\psi\rightarrow \psi+\pi$ translates into the identification $w\rightarrow w+1$ of the annulus, and NS5-branes are smeared along the real line. 
	This is different from the choice in \cite{D'Hoker:2016rdq}, where $w = (\cos\theta)^{2/3}+ i \beta \psi$.} 
\begin{equation}
\label{omega}
w=\frac{1}{\pi}\Bigl(\psi+\frac{i}{\beta}(1-(\cos{\theta})^{2/3})\Bigr), 
\end{equation}
the annulus is depicted in Figure \ref{fig:annulus_ATD}. In (\ref{omega}) we have taken
\begin{equation}
\label{beta}
\beta=\frac{4 m^{1/3}}{3 L^2},
\end{equation}
such that in the supergravity limit $\beta\rightarrow 0$.

As we discussed in section \ref{ATDsub}, the supergravity fields associated to the Abelian T-dual solution describe the near brane geometry of NS5-branes close to the singularity at $\theta=0$ and of D7/O7 branes close to the singularity at $\theta=\pi/2$. This suggests the presence of poles associated to NS5-branes at ${\rm Im}(w)=0$ and of punctures associated to D7/O7 branes at ${\rm Im}(w)=\frac{1}{\beta\pi}$. The fact that the branes are smeared in $\psi$ does however smooth out these singularities from ${\cal A}_\pm$, and leads to the final expressions
\begin{equation}
\label{eq:holo_Ansatz_ATD_new}
{\cal A}_\pm =\frac{3}{8}m\pi^2 \Bigl(w^2-\frac{2i}{\beta\pi}w-\frac{1}{(\beta\pi)^2}\Bigr)
\pm i\frac34 \pi \Bigl(w-\frac{i}{\beta\pi}\Bigr) .
\end{equation}
In the following we show that these expressions emerge from the formalism of DGU for the annulus, taking into account the monodromy of the D7/O7 system as well as the smearing of the branes.

As reviewed in section \ref{annulus} the annulus can be represented in the complex plane by a rectangle with two opposing edges periodically identified. With $w$ defined as in (\ref{omega}), 
the identification $w\rightarrow w+1$ corresponds to the periodicity of the annulus under
$\psi\rightarrow \psi+\pi$. In turn, the periodicity under $w\rightarrow w+\tau$ with $\tau=it$ and $|{\rm Im}(w)|\leq t/2$ is realised with 
\begin{equation}
\label{tau}
\tau=i \frac{2}{\beta\pi}\,, \qquad t= \frac{2}{\beta\pi}\,.
\end{equation}
The surface $\Sigma$ is then defined by
\begin{equation}
\label{Omega}
\Sigma = \Bigl\{ w\in \mathbb{C},\, 0\leq {\rm Re}(w)\leq 1, 0\leq {\rm Im}(w)\leq (\beta\pi)^{-1}\Bigr\}\, ,
\end{equation}
with boundaries
\begin{equation}
\label{boundary}
\partial\Sigma=\Bigl\{ w\in \Sigma, \,{\rm Im}(w)=0 \Bigr\} \cup
\Bigl\{ w\in \Sigma, \,{\rm Im}(w)=\frac{1}{\beta\pi}\Bigr\}.
\end{equation}
As depicted in Figure \ref{fig:annulus_ATD}, there are NS5 branes at the lower boundary, corresponding to $\theta=0$, and D7/O7 branes at the upper boundary, corresponding to $\theta=\pi/2$.

Following section \ref{annulus}, the calculation of the holomorphic functions ${\cal A}_\pm$ for the annulus starts with the computation of the scalar Green function $G$ on $\Sigma$, which vanishes whenever $w\in \partial\Sigma$, and is derived from
\begin{equation}
{\cal G}=\frac{1}{\pi}\int_\Sigma d^2z\, G(w,z)\kappa^2(z), 
\end{equation}
with 
\begin{eqnarray}
&&\partial_{\bar w}\partial_w G(w,z)=-\pi \delta(w,z)\nonumber\\ 
&&G(w,z)|_{w\in \partial\Sigma_1}=0,
\end{eqnarray}
such that $\partial_w \partial_{\bar w}{\cal G}=-\kappa^2$, 
$\kappa^2$ and ${\cal G}$ are the functions introduced in equations (\ref{eq:functions_DHoker}), 
and $\partial\Sigma_1$ denotes the lower boundary of $\Sigma$.
From
\begin{equation}
\lambda (w)\equiv \frac{\partial_w {\cal A}_+}{\partial_w {\cal A}_-} \;,\qquad -\log|\lambda(w)|^2=\sum_{n=1}^N q_n G(w,s_n),
\end{equation}
where $s_n$ are the zeros of $\lambda$, $G$ is guaranteed to also vanish at the upper boundary, i.e.~for $|\lambda|^2=1$. Once the Green function has been obtained, the next step is to construct the holomorphic function $\lambda(w|\tau)$, followed by $\partial_w {\cal A}_\pm$, and finally ${\cal A}_\pm$.
The final expressions for ${\cal A}_\pm$ are then given by equation (\ref{eq:singlevaluedApmannulus}) in section \ref{annulus}. 
As we showed in section \ref{rs7branesannulus}, in the presence of 7-branes, as is the case for the Abelian T-dual solution, these expressions are modified onto, \beq
\label{ApmD7}
{\cal A}_{\pm} ={\cal  A}_{\pm}^0 + \sum_{l=1}^L Y^l_{\pm} \log(\theta_1(w-p_l|\tau))+{\eta}_\pm\int_1^w dz f(z)\sum_{l=1}^L Y^l\partial_{z}\log(\theta_1(z-p_l|\tau)),
\eeq
with $Y^l_{\pm},Y^l$ defined as in \eqref{eq:Ydefs}, $f(z)$ given by equation (\ref{f-annulus}), and where we have adapted expression (\ref{eq:annulusD7}) to account for the anti D7-branes present in the Abelian T-dual solution.

In the remainder of this section we will use this expression to reproduce the holomorphic functions 
${\cal A}_\pm$, given by eq. (\ref{eq:holo_Ansatz_ATD_new}), that define locally the Abelian T-dual solution. We will see that ${\cal A}_\pm$ arise directly from (\ref{ApmD7}) in the supergravity limit $\beta\rightarrow 0$, once the smearing of the NS5 and D7/O7 branes (and thus of the poles in \cite{DHoker:2017mds}) in ${\rm Re}(w)$ is taken into account. It is worth stressing, however, that as a result of the smearing some of the regularity conditions used to derive equation (\ref{ApmD7}) will not be satisfied. For instance, the expressions for $\kappa^2$ and ${\cal G}$ for the Abelian T-dual solution, given by
\begin{equation}\label{eq:ATD_kappa_G}
\kappa^2=i\frac{243}{512}L^6 \beta^3 \pi^3 \left( w-\bar{w}-i\frac{2}{\b\pi} \right)\, , \qquad
{\cal G}=\frac{81}{128}L^6\left[ 1+i\frac{\beta^3 \pi^3}{8}\left(w-{\bar w}-i\frac{2}{\b\pi}\right)^3 \right],
\end{equation}
are related by $\partial_w \partial_{\bar w}{\cal G}=-\kappa^2$, but do not satisfy the regularity conditions \eqref{eq:kappa_G_boundary}. Therefore, they cannot be related through a Green function vanishing in the two boundaries of $\Sigma$.
Furthermore, in the presence of just one zero in $\Sigma$, 
\[s_1= -\frac{i}{m\pi} +\frac{i}{\b\pi} ,\]
as opposed to the derivation in \cite{DHoker:2017mds}, the single-valuedness under $w\rightarrow w+1$ of $\lambda(w|\tau)$, which implies that $\sum_{n=1}^N (s_n - \bar{s}_n)\in \mathbb{Z} \tau$, is also not satisfied. Instead, we have
\begin{equation}
\label{regcon}
s_1-\bar{s}_1=2s_1= -i\frac{2}{m\pi} +i\frac{2}{\b\pi} \notin \mathbb{Z} \left(i \frac{2}{\beta\pi}\right),
\end{equation}
and  $\sum_{n=1}^N (s_n - \bar{s}_n)\in \mathbb{Z} \tau$ is only satisfied in the supergravity limit $\beta\rightarrow 0$, where the first term is negligible.

\medskip
Let us start analysing the contribution of the NS5-branes, given by the term with $\log{\theta_1(w-p_l|\tau)}$ in (\ref{ApmD7}). 

Given that in (\ref{tau}), $\tau\rightarrow i\infty$ in the supergravity limit $\beta\rightarrow 0$, we can take $\tau=it$ and
use the asymptotic expansion for the Jacobi theta-function 
\begin{equation}
\theta_1(z\,|\,\tau) \Big\lvert_{t\to\infty}=2 e^{-\frac{\pi}{4} t}\sin(\pi z)+...,
\end{equation}
for $z= w-p_l$ and $\tau=it=2i/(\beta\pi)$. Using this and taking into account the scaling with $\beta$ of ${\rm Im}(w)$ in eq.~\eqref{omega}, we get that
\begin{equation}
\label{approxomega}
\theta_1(w-p_l |\tau)\approx i e^{-\frac{\pi}{2\beta}}e^{-i\pi (\o-p_l)}.
\end{equation}
Omitting the constant terms, that will finally be absorbed in ${\cal A}^0_\pm$ in (\ref{ApmD7}),
we then get
\begin{equation}
\log{\theta_1(w-p_l|\tau)}\approx -i\pi w.
\end{equation}
The final contribution of the NS5-branes to equation (\ref{ApmD7}) is then
\begin{equation}
{\cal A}_\pm |_{NS5}=-i\pi \Bigl(\sum_{l=1}^L Y_\pm^l\Bigr)\, w .
\end{equation}
Using that $Y^l_\pm=-Z^l_\pm$, since, according to equation (\ref{eq:Page_charges_ATD}), we have anti D7-branes and therefore $p=-1$ and $q=0$, and that
$\sum_{l=1}^L Y_\pm^l=\mp\frac34 \frac{1}{\pi}\int_0^\pi d\psi=\mp\frac34$, due to the smearing of the branes, we find
\begin{equation}
{\cal A}_\pm |_{NS5}=\pm i \frac34 \pi w.
\end{equation}
This exactly reproduces the linear term in $w$ not proportional to the mass (and therefore arising from NS5-branes) in equation (\ref{eq:holo_Ansatz_ATD_new}).

Let us consider now the contribution of the D7/O7 punctures, given by the term 
\begin{equation}
\label{intf(z)}
{\cal A}_\pm |_{D7/O7}={\eta}_\pm\int_1^w dz f(z)\sum_{l=1}^L Y^l\partial_{z}\log(\theta_1(z-p_l|\tau))
\end{equation}
in (\ref{ApmD7}), with $f(z)$,
\beq
f(z) = \sum_{i=1}^I\frac{ n_i^2}{4 \pi}\bigg(\log\left(\gamma_i \frac{\theta_1(z-z_i|\tau)}{\theta_1(z-\overline{z_i}|\tau)}\right)- \frac{2\pi i }{\tau} (z_i-\overline{z_i})z\bigg),
\eeq
as in equation (\ref{f-annulus}). Using the approximation that led to equation (\ref{approxomega}) we find 
\begin{equation}
\frac{\theta_1(z-z_i |\tau)}{\theta_1(z-{\bar z}_i |\tau)}\approx e^{i\pi(z_i-{\bar z}_i)}.
\end{equation}
Using that
\begin{equation}
\partial_{z}\log(\theta_1(z-p_l|\tau))\approx -i\pi,
\end{equation}
as also inferred from (\ref{approxomega}), and that we have anti D7-branes, we have that
\begin{equation}
{\cal A}_\pm |_{D7/O7}=\frac{i\pi}{4} \Bigl(\sum_{l=1}^L Y^l\Bigr)\Bigl(\sum_{i=1}^I n_i^2 (z_i-\bar{z}_i)\Bigr)\Bigl(iw-\frac{\beta\pi^2}{2}w^2\Bigr),
\end{equation}
where we have also omitted constant terms that will be absorbed in ${\cal A}^0_\pm$. Taking into account that $Y^l=Z^l_+-Z^l_-$ and that $Z^l_-=-\overline{Z^l_+}$, and taking 
the D7/O7 branes to lie at $z_i=\psi/\pi +i/(\beta\pi)$ (that is, at $\theta=\pi/2$ and smeared on the $\psi$-direction), and that we can compute $\sum_{i=1}^I n_i^2$ (the sum of the charges of the D7/O7 branes) as $\sum_{i=1}^I n_i^2=m \int_0^\pi d\psi=m\pi$, we finally arrive at
\begin{equation}
{\cal A}_\pm |_{D7/O7}=-\frac34 \frac{m\pi}{\beta}\Bigl(iw-\frac{\beta\pi}{2}w^2\Bigr),
\end{equation}
which exactly reproduces the linear and quadratic terms proportional to the mass in equation (\ref{eq:holo_Ansatz_ATD_new}). 
Adding ${\cal A}_\pm={\cal A}_\pm |_{NS5}+{\cal A}_\pm |_{D7/O7}$ we  reproduce the holomorphic functions given by equation (\ref{eq:holo_Ansatz_ATD_new}) for the Abelian T-dual solution.

We have thus seen that it is possible to find an explicit solution for the annulus in the presence of monodromies for smeared branes, in which case some of the regularity conditions derived in \cite{DHoker:2017mds,DHoker:2017zwj} are not satisfied. The Abelian T-dual of the BO solution provides such a simple explicit example. The smearing of the localised poles and punctures in \cite{DHoker:2017mds,DHoker:2017zwj} does give rise to the two singularities present in the Abelian T-dual solution, associated to the external, smeared, NS5 and D7/O7 branes. 

\section{The non-Abelian T-dual as a DGU solution for the upper half-plane}
\label{sec:NATD_DGU}

The non-Abelian T-dual of the Brandhuber-Oz solution has recently been shown \cite{Hong:2018amk} to fit in the local class of solutions in \cite{D'Hoker:2016rdq}. In this section we show that it also fits globally, with $\Sigma$ the upper half-plane \cite{DHoker:2017mds}.  As in the previous section, it will be necessary to account for the monodromies of 7-branes \cite{DHoker:2017zwj} and explicitly take into account the smearing of the branes.

\begin{figure}
	\centering
	\includegraphics[scale=1.3]{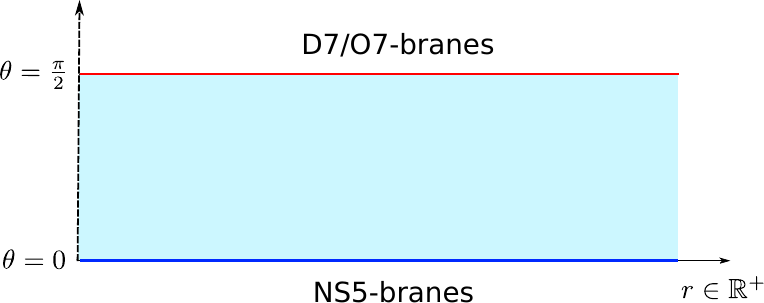}
	\caption{The infinite strip for the non-Abelian T-dual background. NS5-branes are smeared along the lower boundary at $\theta=0$ and D7/O7 branes are smeared along the upper boundary at $\theta=\pi/2$. The strip topology follows from the unboundness of the $r$ direction, $r\in \mathbb{R}^+$.}
	\label{fig:strip_NATD}
\end{figure}

As in the Abelian case, the $(r,\theta)$-directions of the non-Abelian T-dual solution parameterise a two dimensional Riemann surface $\Sigma$ which, given that $r\in \mathbb{R}^+$, has now the topology of an infinite strip. This is depicted in Figure \ref{fig:strip_NATD}.
Taking the same choice of $w$ as in the previous section
\begin{equation}\label{eq:NATD_w_def}
w=\frac{1}{\pi}\Bigl( r+\frac{i}{\beta}\left(1-(\cos{\theta})^{2/3}\right)\Bigr),
\end{equation}
we have NS5-branes in the real line and D7/O7 branes at the upper boundary at ${\rm Im}(w)=\frac{1}{\beta\pi}$, as in the Abelian case. Indeed, the relation with the Abelian T-dual solution suggests that we take NS5 and D7/O7 branes to account for the charges of the external branes, as we will be doing in this section. These branes will also play a role in a possible brane set-up associated to the non-Abelian T-dual solution, discussed in section \ref{field-theory}. Note that in the supergravity limit $\beta\rightarrow 0$ the upper boundary is at infinity. As in the Abelian T-dual  limit, the branes are smeared in $r$, smoothing out the poles and punctures of the holomorphic functions ${\cal A}_\pm$. 

The holomorphic functions that give rise to the non-Abelian T-dual solution are given by
\begin{equation}
\label{ApmNA}
{\cal A}_\pm=\frac18 m\pi^3 \Bigl(w^3-\frac{3i}{\beta\pi}w^2-\frac{3}{(\beta\pi)^2}w+\frac{i}{(\beta\pi)^3}\Bigr)\pm i\frac34 \pi \Bigl(w-\frac{i}{\beta\pi}\Bigr)+i \frac{m}{4\beta^3}.
\end{equation} 
In the following we show that we can reproduce the finite terms in this expression, in the limit $\beta\rightarrow 0$, using  the formalism of DGU for the upper half-plane \cite{DHoker:2017mds}, taking also into account the monodromies from 7-branes  \cite{DHoker:2017zwj}.

We recall from section \ref{inclusion7branes}  the general expressions for ${\cal A}_\pm$ for the upper half-plane in the presence of monodromies
\begin{equation}
\label{ApmD7bis}
{\cal A}_\pm={\cal A}^0_\pm+\sum_{l=1}^L Y_\pm^l\log{(w-p_l)}+ \eta_\pm\int_{\infty}^w dz f(z) \sum_{l=1}^L \frac{Y^l}{z-p_l},
\end{equation}
where $f$ is given by (\ref{fomega}) and $Y^l= Z^l_+-Z^l_-$. As we have anti D7-branes, we set $p=-1$ and $q=0$, so that $Y_\pm^l=-Z_\pm^l$.

Let us start analysing the contribution of the NS5-branes, given by the term with $\log{(w-p_l)}$ in (\ref{ApmD7bis}). 
As discussed in section \ref{NATD}, in the non-Abelian T-dual solution a NS5-brane is created 
each time an $r_p=p\pi$ value in the $r$ direction is crossed. This charge is however smeared in this direction, as it happens to the
charges associated to the D7/O7 branes. Moreover, given the unboundedness of the $r$ direction, the number of branes is infinite. Thus, we will have to
conveniently regularise the infinite sum in $l$ that appears in equation (\ref{ApmD7bis}) such that we can recover a finite contribution to ${\cal A}_\pm$ in the $r\rightarrow\infty$ limit.

It turns out that a convenient way to regularise is to take NS5-branes at $\frac{r}{\pi}-i\alpha$ for non zero $\alpha$, smear them in $r\in [0,n\pi]$ and compensate for the charge on the imaginary axis with their images at $\frac{r}{\pi}+i\alpha$. Tuning $\alpha$ we can then reproduce the $\pm i\frac34\pi w$ contribution from NS5-branes to ${\cal A}_\pm$.

Taking into account the NS5-brane charge density in the $[0,n\pi]$ interval, given by equation (\ref{NScharge}), we have
\begin{eqnarray}
\sum_l Z^l_\pm  \log{(w-p_l)}&=&\mp \frac{3}{8}\int_0^{n\pi}\frac{dr}{\pi} \log{\Bigl(\frac{w-\frac{r}{\pi}-i\alpha}{w-\frac{r}{\pi}+i\alpha}\Bigr)}= \pm \frac{3i}{4\pi}\left\{ \frac{n\pi^2}{2}+ \int_{0}^{n\pi} dr \arctan{\left(\left(w-\frac{r}{\pi}\right)\frac{1}{\alpha}\right)} \right\}\nonumber\\
&&
\end{eqnarray}
where we have used that
\begin{equation}
\log{\left((-1)\frac{w-i\alpha}{w+i\alpha}\right)}=\log{\left(\frac{1+i\o/\a}{1-i\o/\a}\right)}=2i\, \arctan{\frac{w}{\alpha}}+i\pi.
\end{equation}
Taking in this expression the double limit $n\rightarrow \infty$ and $\alpha\rightarrow \infty$ such that $n/\alpha$ is kept finite, and keeping only the finite terms, we arrive at the following expression, \begin{equation}
\sum_l Z^l_\pm  \log{(w-p_l)}\approx\mp i\frac34 w \Bigl(\pi - \arctan{\frac{n}{\alpha}}\Bigr).
\end{equation}
Taking into account that $Y^l_\pm=-Z^l_\pm$ we thus reproduce the finite contribution $\pm i\frac34 \pi w$ to ${\cal A}_\pm$ in the limit $n/\alpha\rightarrow 0$. 

A similar calculation gives rise to the cubic contribution to ${\cal A}_\pm$. In this case we focus on the last term in equation (\ref{ApmD7bis}). Here we first calculate the regularised expression of the sum $\sum_{l=1}^L \frac{Y^l}{z-p_l}$ as the derivative with respect to $z$ of our previous approximation to $\sum_{l=1}^L Y^l  \log{(z-p_l)}$. This gives
\begin{equation}
\label{suminl}
\sum_{l=1}^L \frac{Y^l}{z-p_l} \approx  i\frac32 \Bigl(\pi - \arctan{\frac{n}{\alpha}}\Bigr)\approx   i\frac32 \pi,
\end{equation}
where we have taken into account that $Y^l=Z^l_+-Z^l_-$ and that $Z^l_-=-\overline{Z^l_+}$.
Next, we compute the function $f(z)$, given by equation (\ref{fomega}). In this case we have D7/O7 branes located at $z_i=\frac{r}{\pi}+\frac{i}{\beta\pi}$, smeared in $r\in [0,n\pi]$, with a charge density given by $dN_{7}^r=-mrdr$, according to equation (\ref{ND7r}). Substituting in $f(z)$ we find\\
\begin{equation}
f(z) 
=-\frac{m}{4\pi} \int_{0}^{n\pi}dr\, r \log{\left(\frac{z-\frac{r}{\pi}-\frac{i}{\beta\pi}}{z-\frac{r}{\pi}+\frac{i}{\beta\pi}}\right)}=-\frac{i}{2\pi}m\int_{0}^{n\pi}dr\, r \left[ \frac\pi2 + \arctan{\left(\left(z-\frac{r}{\pi}\right)\beta\pi\right)} \right] .
\end{equation}
Taking now the double limit $n\rightarrow \infty$, $\beta\rightarrow 0$ such that $n\beta$ is finite and small, and keeping only finite terms, we find that
\begin{equation}
\label{f(z)}
f(z)\approx i\frac{m\pi}{4}z^2 \Bigl(\pi-\arctan{(\pi n \beta)}\Bigr)\approx  i\frac{m\pi^2}{4}z^2.
\end{equation}
Substituting (\ref{f(z)}) and (\ref{suminl}) in equation (\ref{ApmD7bis}) and integrating in $z$ we finally reproduce the $\frac18 m\pi^3w^3$ cubic contribution to ${\cal A}_\pm$ in (\ref{ApmNA}). 

Adding the previous contributions of NS5 and D7/O7 branes we finally get
\begin{equation}
{\cal A}_\pm ={\cal A}_\pm^0 +\frac18 m\pi^3w^3\pm i\frac34 \pi w.
\end{equation}
These expressions fit the finite terms in $w^3$ and $w$ present in equation (\ref{ApmNA}). 
Our regularisation scheme does not however allow us to reproduce the infinite, in the $\beta\rightarrow 0$ limit, yet, $w$-dependent, terms in (\ref{ApmNA}), which cannot be either reproduced tuning the constants ${\cal A}^0_\pm$ as we did in the Abelian case. Unfortunately we have not been able to come up with the right regularisation scheme that allows us to recover also such terms.

Still, a particularly interesting output of our analysis is that we can identify the precise way in which $r$ grows to infinity in the supergravity limit in order to generate a finite solution. Namely, $r\rightarrow\infty$ such that $r<< L^2$, as inferred from equations (\ref{beta}) and (\ref{eq:NATD_charges}). This is likely to be extrapolated to other non-Abelian T-dual solutions in which the competition between these two limits remained unclear (see for instance section 5 in \cite{Lozano:2016wrs}).
We will discuss the implications of this scaling in section \ref{field-theory} when we address the CFT interpretation of the non-Abelian T-dual solution.

\section{Holographic central charge and entanglement entropy}
\label{entanglement}

In this section we compute the holographic central charge and entanglement entropy of the $AdS_6$ T-duals of the Brandhuber-Oz solution. The central charge is derived directly from the Abelian T-dual and non-Abelian T-dual solutions, while the entanglement entropy is derived using their realisations as DGKU solutions. These observables are expected to exhibit the same scaling with the quantised charges, and, in the non-Abelian T-dual case, can provide useful input about the dual SCFT\footnote{The entanglement entropy for the non-Abelian T-dual solution restricted to the $r\in [0,\pi]$ region was computed in \cite{Lozano:2013oma}.}.

\subsection{Holographic central charge}

The holographic central charge is computed using the formalism developed in \cite{Klebanov:2007ws},\!\!\!\cite{Macpherson:2014eza}, with the particular conventions of \cite{Itsios:2017cew}. Given a metric and dilaton of the form,
\begin{equation}
ds^2=a(r_0,\theta^i) \, \Big[dx_{1,d}^2 +  b(r_0) \, dr^2\Big] + g_{ij}(r_0,\theta^i) \, d\theta^i \, d\theta^j,  \;\;\; \Phi(r_0,\theta^i),
\label{xxy}
\end{equation}
where $r_0$ denotes the AdS radial direction and the $\theta^i$ the internal space directions, we define: 
\begin{equation}
\label{cc2}
\hat{V}_{int}=\int\! d\theta^i\sqrt{\det [g_{ij}] \, e^{-4\Phi} \, a^{d}}\,,\qquad \hat{H}=\hat{V}_{int}^2\,.
\end{equation}
This specifies the holographic central charge of the $(d+1)$-dimensional QFT as follows,
\beq
c=\pi \, d^d \, \frac{b^{d/2 } \hat{H}^{\frac{2d+1}{2}}}{G_{N,10} \, \big( \hat{H}' \big)^d}\,,   \;\;\;\;\; G_{N,10}= 8 \, \pi^6 g_s^2 \, \alpha'^4 .
\label{formulacentralcharge}
\eeq

Particularising to the Brandhuber-Oz background (we set $\a'=g_s=1$):
\beq
a=\frac{9}{4}L^2r_0^2 W^2,\;\; b=\frac{1}{r_0^4}, \;\; d=4, \;\; \sqrt{e^{-4\Phi} \, \det[g_{ij}] \, a^3}= \frac{729}{512} L^{10} r_0^4 \sin^3\theta \, \sin\theta_1\,W^{-2} \, ,
\eeq
and, finally:
\begin{equation}
\label{ccBO}
c = \frac{27}{40\sqrt{2}} \frac{N_{D4}^{5/2}}{N_{8}^{1/2}} \,,
\end{equation}
where the quantised charges are those introduced in eq.~\eqref{eq:BO_charges}.

Similarly, the central charge for the Abelian T-dual solution is given by
\begin{equation}
\label{cATD}
c_{ATD} = \frac{27}{40\sqrt{2}} \frac{N_{D5}^{5/2}}{N_{7}^{1/2}} \,,
\end{equation}
where the charges are those in eq.~\eqref{eq:ATD_charges}. One thus gets the same central charge of the BO solution, with $N_{D4}$ replaced by $N_{D5}$ and $N_8$ by $N_7$, in agreement with the fact  that the Abelian T-dual solution describes the same CFT as the BO solution, but realised in a (D5, D7/O7, NS5) brane system. We will see the same result arising from the computation of the entanglement entropy, using the explicit realisation of the Abelian T-dual background as a solution in the class of \cite{D'Hoker:2016rdq}.

The same computation in the non-Abelian T-dual case yields an infinite central charge unless the radial direction is restricted to a finite interval. Taking $r$ to lie in $r\in[0,n\pi]$ we obtain
\begin{equation}\label{cNATD}
c_{NATD} = \frac{9}{40\sqrt{2}\pi} n^3\, \frac{N_{D7}^{5/2}}{m^{1/2}} \,,
\end{equation}
where $N_{D7}$ is given by equation (\ref{eq:Page_charges_NATD}). We choose these parameters to express the central charge in the $r\in[0,n\pi]$ interval because they take the same values in all $[p\pi,(p+1)\pi]$ intervals for $p=0,\dots,n-1$.

In turn, the expression of the central charge in each  $r\in [p\pi,(p+1)\pi]$ interval is given by
\begin{equation}\label{cNATDn}
c_{NATD}^{(p,p+1)} = \frac{9}{40\sqrt{2}\pi}(3p^2+3p+1) \frac{N_{D7}^{5/2}}{m^{1/2}} \,.
\end{equation}
For large $p$ this becomes
\begin{equation}
c_{NATD}^{(p,p+1)} = \frac{27}{40\sqrt{2}} \frac{N_{D5}^{5/2}}{(N_7^r)^{1/2}} \,,
\end{equation}
where, given that we are referring to a $[p\pi,(p+1)\pi]$ interval, we can use the D5 and 7-brane conserved charges $N_{D5}$ and $N_7^r$ referred to that interval. This expression is
in full agreement with the expression \eqref{cATD} for the Abelian T-dual solution, as expected for pairs of Abelian and non-Abelian T-duals (see \cite{Lozano:2016kum}). 

Finally, note that the central charge in the full $r\in [0,n\pi]$ interval can be simply obtained as
\begin{equation}
c_{NATD}=\sum_{p=0}^{n-1} c_{NATD}^{(p,p+1)}
\end{equation}

\subsection{Entanglement entropy}

We now compute the holographic entanglement entropy, as derived by Ryu and Takayanagi \cite{Ryu:2006ef,Casini:2011kv}, for the 5d SCFTs dual to the Abelian and non-Abelian T-dual solutions.
The match between the entanglement entropy for a spherical entangling surface and the free energy\footnote{For an odd-dimensional SCFT, the finite contribution to the entanglement entropy of a ball-shaped region is expected to agree with the renormalised free energy, up to a sign \cite{Casini:2011kv}.} of the field theory on $S^5$ \cite{Jafferis:2012iv} was exploited in \cite{Gutperle:2017tjo} to give further support to the interpretation of the DGKU/DGU solutions as gravity duals of 5d SCFTs living on 5-brane webs.

The entanglement entropy is given by the area of the entangling surface of the SCFT, which is a codimension-2 surface of the $AdS_6$ space, anchored at a fixed time on its boundary. The Ryu-Takayanagi \cite{Ryu:2006ef} prescription reads
\begin{equation}\label{eq:SEE_def}
S_{EE} = \frac{\textrm{Area}(\g_8)}{4G_N} ,
\end{equation}
where $\textrm{Area}(\g_8)$ is the codimension two minimal area coupled to the entangling surface at the boundary. Following reference \cite{Gutperle:2018vdd}, where the entanglement entropy for 
$AdS_6  \times S^2 \times \Sigma$ solutions including 7-branes was computed, we have that:
\begin{align}\label{eq:Area8}
\textrm{Area}(\g_8) &= \int_{\g_8} (\l_6^4\,\textrm{vol}(\g_4)) \wedge (\l_2^2\,\textrm{vol}(S^2))\wedge (\textrm{vol}(\Sigma)) \\[2mm] 
& = \int_{\g_4} \textrm{vol}(\g_4) \int_{S^2} \textrm{vol}(S^2) \int_{\Sigma} \!\!\l_6^4\, \l_2^2\,\textrm{vol}(\Sigma) = 4\pi\,\textrm{Area}_{ren}(\g_4) \,\mathcal{J}. \nn
\end{align}
Here $\lambda_6^2$ and $\lambda_2^2$ are defined as in equation (\ref{eq:metric_DHoker}),
$\textrm{Area}_{ren}(\g_4)$ is the finite area of the 4-dimensional ball-shaped surface inside the unit radius $AdS_6$ space,
\[\textrm{Area}_{ren}(\g_4)= \frac{2}{3} \,\textrm{Vol}(S^3) =\frac{4\pi^2}{3}, \]
and $\mathcal{J}$ encodes the integral over the Riemann surface $\Sigma$ which, considering the expressions for the metric factors in eq.~\eqref{eq:factors_DHoker}, can be written as:
\begin{align}
\mathcal{J} &= \int_{\Sigma}\!\! \l_6^4\,\l_2^2\, \textrm{vol}(\Sigma)
= 4 \int_{\Sigma}\!\!\!\!d^2w\; \l_6^4\,\l_2^2\,\tilde{\rho}^2
= \frac{8}{3} \int_{\Sigma} \!\!\!\!d^2w\; \kappa^2 \mathcal{G}
= \frac{8}{3} \int_{\Sigma} \!\!\!\!d^2w\; |\partial_w\mathcal{G}|^2 \,,
\end{align}
where in the last step a partial integration has been performed after the substitution $\kappa^2= -\partial_{\bar w}\partial_{w} \mathcal{G}$\footnote{This partial integration gives rise to a boundary term of the form $\int d\bar{w}\,\mathcal{G} \partial_{\bar w}\mathcal{G}$, that either vanishes identically for regular solutions or gives rise to subleading contributions in the supergravity limit.}.
The final expression for the entanglement entropy of a DGKU solution is then given by:
\begin{equation} 
S_{EE}=\frac{32\pi^3}{9G_N} \!\!\int_{\Sigma} \!\!\!\!d^2w\; |\partial_w\mathcal{G}|^2 .
\end{equation}


We next particularise this to the Abelian and non-Abelian T-dual solutions. Using the parameterisation of $w$ given in eq.~\eqref{omega} and the $\mathcal{G}$ function in \eqref{eq:ATD_kappa_G} we find, for the Abelian T-dual solution
\begin{equation} 
\label{EEATD}
S_{EE}=\frac{32\pi^3}{9G_N} \b \int_0^1 \!\!dx\int_0^\pi \!\!d\psi\; \frac{81}{16} \frac{m^2x^4}{\b^6} = \frac{18\pi}{5\sqrt{2}}\frac{N_{D5}^{5/2}}{N_7^{1/2}} \,.
\end{equation}
As expected, this exactly matches the entanglement entropy of the BO solution, computed in \cite{Jafferis:2012iv}.
In turn, we find the following finite contribution to the entanglement entropy of the non-Abelian T-dual solution, for $r\in [0,n\pi]$, 
\begin{align} 
\label{EEnpi}
S_{EE}&=\frac{32\pi^3}{9G_N} \b \int_0^1 \!\!dx\int_0^{n\pi} dr \; \frac{9m^2}{16\b^8} \left[ (1-x^3)^2+9\b^2x^4r^2 \right] \approx \frac{6}{5\sqrt{2}}\,n^3\, \frac{N_{D7}^{5/2}}{m^{1/2}}  
.
\end{align}
For $r\in [p\pi,(p+1)\pi]$ we find
\begin{equation}
S_{EE}^{(p,p+1)}=\frac{6}{5\sqrt{2}}(3p^2+3p+1)\frac{N_{D7}^{5/2}}{m^{1/2}} .
\end{equation}
From here we can recover (\ref{EEnpi}) summing up from $p=0$ to $n-1$. We can also recover eq.  (\ref{EEATD}) for the Abelian T-dual solution in the large $p$ limit.



As expected, the entanglement entropies coincide with the holographic central charges up to a constant. Both observables confirm the $N^{5/2}$ scaling with the colour charge expected for $USp(2N)$ gauge theories \cite{Jafferis:2012iv}. Note that this is intrinsically different from the $N^4$ behaviour found in \cite{Gutperle:2018vdd} for the $(p,q)$ 5-brane webs considered therein.

Regarding $n$, it is interesting to note that both the central charge and entanglement entropy of the non-Abelian T-dual solution scale with $p^2$ at large $p$, for $r\in [p\pi,(p+1)\pi]$, such that the values of these observables for the BO solution are reproduced. In turn, this scaling is with $n^3$ for $r\in [0,n\pi]$. This would be compatible with a field theory built up of $USp(pN_{D7})$ gauge groups at each $r\in [p\pi,(p+1)\pi]$ interval.
This would explain the $n^3$ scaling of the central charge and entanglement entropy as the result of adding the $p^2$ contributions of the $USp$ theories with increasing ranks living in each interval. Note however that the field theory interpretation of these observables, computed at each, length $\pi$, interval would require further study \footnote{We would like to thank Diego Rodr\'{\i}guez-G\'omez for an interesting discussion on this point.}.

\section{On the CFT interpretation of the non-Abelian T-dual solution}
\label{field-theory}

The global $AdS_6$ solutions constructed in  \cite{DHoker:2017mds,DHoker:2017zwj} describe holographically 5d CFTs realised in $(p,q)$ 5-brane webs  \cite{Aharony:1997ju,Aharony:1997bh,DeWolfe:1999hj} in Type IIB string theory \cite{DHoker:2017mds,Gutperle:2017tjo,Gutperle:2018vdd,Bergman:2018hin,Gutperle:2018wuk,Fluder:2018chf,Kaidi:2017bmd,Kaidi:2018zkx}. Interesting specific examples have been discussed  in these references, and quantitative matches between field theory and holographic results have been obtained.
In this section we speculate with a possible 5-brane web compatible with the charges of the non-Abelian T-dual solution. 

The analysis of charges performed in section \ref{NATD} suggests that we are dealing with D5-branes extended on $(\mathbb{R}^{1,4},r)$, and NS5-branes extended on $AdS_6$
and localised at $r_p=p\pi$. At each interval $[r_p,r_{p+1}]$ there are $pN_{D7}$ D5-branes stretched between the NS5-branes at both ends.  On top of this,
there are D7/O7 flavour branes whose charges at each interval are given by equation (\ref{ND7r}).

Omitting for the moment the contribution of the D7/O7 brane system, the analogy with the Abelian T-dual case in the $r\rightarrow \infty$ limit suggests a brane set-up in which:
\begin{enumerate}
	\item $x^9$ parameterises the Coulomb branch of the moduli space, while $r$ is the field theory direction. Note that since $r\in \mathbb{R}^+$ the brane set-up would be infinite.
	\item In each $r\in [p\pi,(p+1)\pi]$ interval with $p=1,2,\dots$ there are $p N_{D7}$ D5-branes stretched between NS5-branes at $r_p=p\pi$ and at $r_{p+1}=(p+1)\pi$. The NS5-branes are bent due to the different number of D5-branes ending on each side.
	\item The brane set-up is symmetric in $x^9$. This is inherited from the $\mathbb{Z}_2$ action of the O8 fixed plane of the BO solution.
\end{enumerate}


In this 5-brane web the Coulomb branch would correspond to the positions of the D5-branes in the $x^9$ direction. In turn, the lengths of the stretched D5-branes would correspond to the inverse of the gauge couplings, 
with the bare couplings arising at the origin of the Coulomb branch, where the 5-brane web collapses and only the external 5-branes remain.

Regarding the contribution of the D7/O7 brane system,
the worldsheet analysis performed in Appendix \ref{worldsheet} shows that the O8 fixed plane of the BO solution transforms onto a O5 fixed plane located at $r=0$, with the non-Abelian T-dual  solution describing the $r\ge 0$ physical region. Note that at $r=0$ the O5 can be reinterpreted as a O7 wrapped on $S^2$, with the $S^2$ collapsed to a point, consistently with our argumentation in section \ref{NATD} for an underlying (D5, NS5, D7/O7) brane system. In the Abelian T-dual limit, in which $r$ is compactified into a $[0,\pi]$ interval at infinity, there is a second O7 fixed plane at $r=\psi=\pi/2$, which is however missing in the uncompactified case. 

\begin{figure}
	\centering
	\includegraphics[scale=0.25]{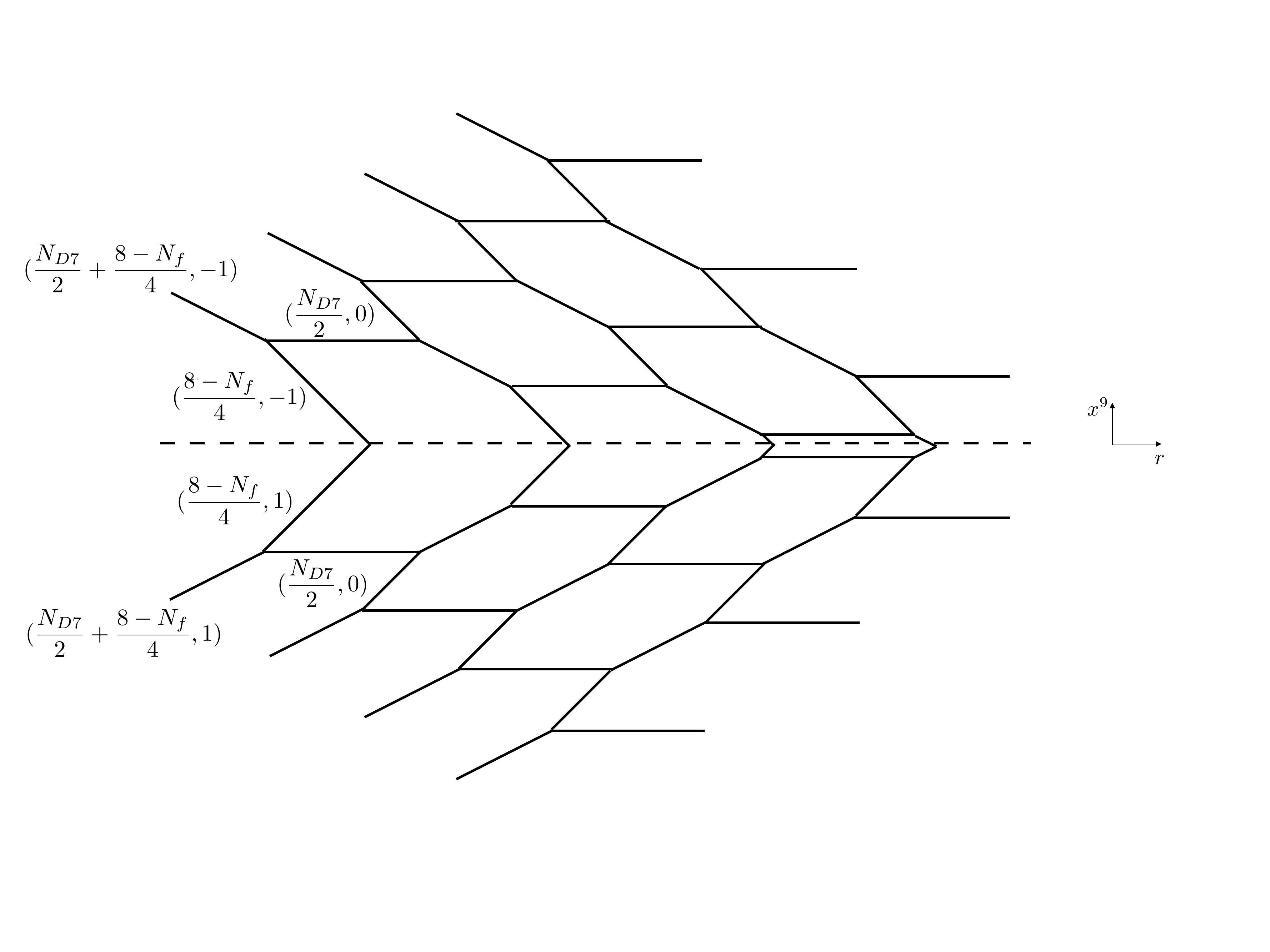}
	\vspace{-1.5cm}
	\caption{5-brane web consistent with the quantised charges of the non-Abelian T-dual solution. The dashed line represents the branch-cut created by the D7/O7 branes at each $r\in [p\pi,(p+1)\pi]$ interval. At each interval a new, smeared, D7/O7 brane system adds to the discontinuity in the 5-brane charge created by the branch cut.}
	\label{extended-set-up}
\end{figure}

Consistently with the previous discussion, we would need to add a 
unique O7 fixed plane \cite{Brunner:1997gk,Bergman:2015dpa} orthogonal to the 5-branes, located at $r=0$, $x^9=0$. The way this can be made compatible with the analysis of the conserved charges is if the fixed plane is smeared in $r$, such that it can give rise,
together with the flavour D7-branes orthogonal to the 5-brane web, to the net $N_7^r=m\pi^2(p+1/2)$ charge at each $r\in [p\pi,(p+1)\pi]$ interval. 
A possible brane set-up compatible with these observations would be the one shown in Figure \ref{extended-set-up}. 

It is interesting to notice that even if the non-Abelian T-dual solution does not seem to capture many of the global symmetries associated to this brane configuration, the previous brane set-up is  suggestive of a linear quiver consisting of infinite gauge groups of increasing ranks, as discussed in the previous section. This could be seen as a generalisation of the 5-brane webs in the presence of O7 fixed planes, describing $USp(2N)$ fixed point theories with $N_F\leq 2N+4$ flavours, discussed in \cite{Bergman:2015dpa}.
Indeed, one can check that at each interval $r\in [p\pi,(p+1)\pi]$, the colour charge $N_{D5}^{(p)}=p N_{D7}$, and the flavour charge $N_{7}^{r(p)}=m\pi^2 p=\frac{8-N_f}{2}\pi p$ (plus their mirrors)\footnote{Here, as compared with equation (\ref{ND7r}), we have shifted all charges by $m\pi^2/2$, and interpreted this overall charge as a background charge decoupled from the brane configuration. The same reasoning led to a consistent Gaiotto-Witten brane set-up \cite{Gaiotto:2008ak} for the $AdS_4$ solution considered in  \cite{Lozano:2016wrs}.} satisfy the condition 
\begin{equation}
\label{condicion}
N_F\leq 2N+4,
\end{equation}
for the existence of a $USp(2N)$ fixed point theory in the UV, in the absence of antisymmetric hypermultiplets\footnote{The absence of antisymmetric hypermultiplets is in agreement with the global symmetries of the non-Abelian T-dual solution, discussed in section \ref{NATD}.} \cite{Intriligator:1997pq,Bergman:2015dpa}. This is seen by splitting the flavour charge into the $4\pi p$ contribution of the (smeared) O7 plane, and the $\frac{N_f}{2}\pi p$ contribution of the D7 flavour branes\footnote{Here the $\pi$ factor has to do with the different normalisation of the brane charges after the non-Abelian T-duality transformation (see section \ref{NATD}).}, and taking into account the bifundamental matter coming from the open strings connecting D5-branes across the two (bent) NS5-branes limiting the interval. This produces a total flavour charge
$N_F=\frac{N_f}{2}\pi p+(p-1)\frac{N_{D7}}{2}+(p+1)\frac{N_{D7}}{2}$, with $N_f<8$. Together with the number of colours in the $p$-th interval, $N=p\frac{N_{D7}}{2}$, it is easy to see that this satisfies the condition (\ref{condicion}). Note that this is the condition found in 
\cite{Intriligator:1997pq} for stringy probe set-ups, as those discussed in this paper.

We should stress that, as in other examples of $AdS$ backgrounds constructed through non-Abelian T-duality \cite{Lozano:2016kum,Lozano:2016wrs,Lozano:2017ole,Itsios:2017cew}, the fact that the field theory direction is non-compact renders an infinite brane set-up, that needs to be completed such that it can describe a well-defined CFT. Inspired by the previous examples, one possible way to terminate the 5-brane web in this case would be to take $r\in [0,n\pi]$ and to add D7-branes (or semi-infinite D5-branes) to the right of the (bent) NS5-brane located at $r=n\pi$. This would ensure that the condition (\ref{condicion}) is satisfied also in the last interval. This completion of the quiver with a hard cut-off would give rise to a {\it completion} of the non-Abelian T-dual geometry that could realise the global symmetries of the quiver not captured by the non-Abelian T-dual solution.

In our current set-up this would mean identifying the effect that the addition of the external $n N_{D7}$ flavour D5-branes has in the geometry. In other examples, such as the $AdS_5\times S^2$ Gaiotto-Maldacena geometry studied in \cite{Lozano:2016kum} and the $AdS_4\times S^2\times S^2$ example considered in \cite{Lozano:2016wrs}, this had the effect of, not only rendering the geometry compact, but also removing the singularities, with the non-Abelian T-dual geometry arising as an effective description far from the added external branes.
It would be interesting to perform a careful analysis along those lines in this case.

Finally, we would like to stress that, as in previous examples, the completion of the field theory, and thereof of the geometry, is not unique. In this example we could for instance have completed the brane set-up by adding a second orientifold fixed plane and the corresponding mirror image of the current brane set-up. It would be interesting to clarify the relation between these different  possibilities for completing non-Abelian T-dual geometries.

\section{Conclusions}
\label{conclusions}

In this paper we have revisited the classification of $AdS_6\times S^2$ global solutions to Type IIB supergravity  in  \cite{DHoker:2017mds,DHoker:2017zwj} to describe solutions arising from delocalised 5 and 7-branes. We have seen that both the Abelian and non-Abelian T-duals of the Brandhuber-Oz solution take the form of $AdS_6\times S^2$ geometries warped over a two dimensional Riemann surface with two boundaries. In the Abelian T-dual case the Riemann surface is an annulus, while in the non-Abelian case it is an infinite strip.  In both cases there are smeared NS5-branes at one of the boundaries and smeared D7/O7 branes at the other boundary, that generalise the localised poles and punctures of the solutions in \cite{DHoker:2017mds,DHoker:2017zwj}. The explicit realisation of the Abelian T-dual solution required extending the results for the annulus in  \cite{DHoker:2017mds}  to include seven branes. 

In addition to recovering the T-dual, another reason to study the annulus with 7-branes is that they may actually be required to have a solution on the annulus at all. In \cite{DHoker:2017mds} a numerical study was performed on the annulus with localised 5-branes only, but was unable to satisfy the regularity conditions on both boundaries simultaneously. Although this does not  constitute a proof, it is suggestive that seven branes are required- we hope to study the annulus with localised seven branes in the future. 
More recently, the authors of \cite{Kaidi:2018zkx}  have given an additional argument for the non-existence of solutions for the annulus. They have claimed that IIB solutions arising from 5-brane webs with more than one boundary, or higher genus, would lead, upon uplift, to curves in M-theory associated to mass deformations of 5-brane webs, which would correspond to renormalisation group flows rather than fixed points. This argument may seem to contradict our findings in this paper, however the set-up in M-theory describes a system of localised branes - ours are smeared which is likely how we side step this issue. Another way out could be that the identification between the Riemann surface and the M-theory curve put forward in \cite{Kaidi:2018zkx} does not hold for Riemann surfaces with non-trivial topologies. It would be interesting to investigate this further.

An interesting general feature of our current analysis of the annulus with 7-branes is that one appears to be restricted to adding $(1,0)$ 7-branes only- ie standard D7's. The reason is that one needs to impose periodicity under $w\to w+1$, and this can only be achieved by making the holomorphic functions at $\text{Re}\,w= 1$ and $\text{Re}\,w= 0$ related by a specific $SL(2,\mathbb{R})$ transformation that leaves the dilaton and gauge invariant fluxes unchanged. Such a transformation is only possible if $q=0$, so we are dealing with D7 branes only. Given a solution on the annulus with D7's one might then ask what the result of mapping this to a $(p,q)$ 7-brane system with another $SL(2,\mathbb{R})$ transformation would be. The axion and dilaton would mix under such a generic transformation, and since the specific $SL(2,\mathbb{R})$ mapping of the holomorphic functions at $\text{Re\,}w= 1$ to  $\text{Re}\,w= 0$  actually shifts the axion by a constant, the dilaton in the $(p,q)$ 7-brane system would no longer be $w\to w+1$ periodic. This feature already exists in the T-dual solution, as $F_1$ has a leg in the periodic direction. The only possible way round this issue is if one can justify performing a gauge transformation on the axion of the D7 brane system as one crosses $\text{Re}\,w= 1$ which cancels the effect of this constant shift - at present such a mechanism eludes us.

We have discussed a possible 5-brane web compatible with the conserved charges of the non-Abelian T-dual solution. This required elucidating the way the O8 fixed plane of the BO solution is transformed under non-Abelian T-duality. Already in simpler brane configurations not involving orientifolds, the main problem in building up brane set-ups associated to non-Abelian T-dual solutions is that it is not known how the brane configurations underlying the original background transform. For this reason one has to resort to the solution itself, after the (would be) near brane limit has been taken, in order to elucidate the properties of the dual CFT.


We have derived through a worldsheet analysis  the explicit non-local mapping that underlies non-Abelian T-duality transformations with respect to the left or right action of a $G_L\times G_R$ isometry group. Using this mapping we have identified\footnote{See also \cite{Lozano:2013oma} for some partial  previous results.} the way parity reversal is transformed under non-Abelian T-duality with respect to a freely acting $SU(2)$: It maps $\Omega$ onto $I_\chi \Omega$, where $I_\chi$ is a reflection through the origin of the $\mathbb{R}^3$ space spanned by the dual variables. This generalises to the non-Abelian case the well-known mapping $\Omega\rightarrow I_9\Omega$, under Abelian T-duality along a compact direction $x^9$. Using this mapping we have seen that the O8 orientifold fixed plane of the BO solution is mapped onto a O5 fixed plane located at the origin of the $\mathbb{R}^3$ dual space, where, as we have discussed, it is indistinguishable from a O7 fixed plane. With this information, and under certain assumptions, based on the agreement with the Abelian T-dual solution in the large $r$ limit and the similarities with previous examples of non-Abelian T-dual backgrounds \cite{Lozano:2016kum,Lozano:2016wrs,Itsios:2017cew}, we have discussed a possible 5-brane web with additional D7-branes and a (smeared) O7 fixed plane compatible with the quantised charges of the solution. 


Our discussion about the CFT interpretation of the non-Abelian T-dual solution leaves interesting open problems to study:
\vspace{-2mm}
\begin{itemize}
	\setlength\itemsep{-0.3em}
	\item Computing the backreaction of the external D7-branes used to complete the infinite linear quiver inferred from the non-Abelian T-dual solution would provide a geometrical completion of this solution that could account for the realisation of properties of the quiver, such as the existence of global symmetries, not captured by the non-Abelian T-dual solution. It should be possible to find out the explicit completed geometry by identifying the holomorphic functions associated to the solution with the extra D7-branes.
	\item Related to the previous point, it would be interesting to investigate in more detail the fixed point theory that arises from a linear quiver of $USp(2N)$ gauge groups with increasing ranks, and its possible relation with the {\it completed} non-Abelian T-dual solution. Further, it would be interesting to explore the global symmetry at the fixed points not visible in supergravity \cite{Bergman:2012kr}, and its possible enhancement, along the lines of \cite{Bergman:2013aca,Bergman:2015dpa}.
	\item An interesting output of our study of the realisation of the non-Abelian T-dual solution as a DGU geometry is that, in contrast with other linear quiver CFTs \cite{Gaiotto:2014lca,Cremonesi:2015bld}, the growth to infinity of the number of nodes of the linear quiver associated to the non-Abelian T-dual solution would be subleading compared to the growth to infinity of the individual ranks of the gauge groups. This is likely to be extrapolated to other non-Abelian T-dual solutions associated to infinite linear quivers \cite{Lozano:2016kum,Lozano:2016wrs,Itsios:2017cew}, where the competition between these two limits remained unclear.
	\item As we have discussed, the regularisation procedure yielding a well-defined dual CFT for the non-Abelian T-dual solution is not unique. It would be interesting to provide a consistent framework that singles out one such possible regularisation\footnote{For instance in the spirit of \cite{Itsios:2017cew}, that used the connection between $\mathcal{N}=2$ and $\mathcal{N}=1$ 4d CFTs through mass deformations.}.
\end{itemize}

\subsection*{Acknowledgements}

We would like to thank Thiago Araujo, Emmanuel Malek, Carlos N\'u\~nez, Eoin \'O Colg\'ain, Christoph Uhlemann and especially Diego Rodr\'{\i}guez-G\'omez for very useful discussions. Y.L.~and J.M.~are partially supported by the Spanish Government Research Grant FPA2015-63667-P. J.M.~was also supported in part by the FPI scholarship BES-2013-064815 of the same institution. N.T.M.~is funded by the Italian Ministry of Education, Universities and Research under the Prin project ``Non Perturbative Aspects of Gauge Theories and Strings'' (2015MP2CX4) and INFN.

\appendix

\section{Details of the annulus in general}
\label{Annulusappendix}
In this appendix we give some further details of our computations for the annulus. In the main text we have assumed that all poles are located on the real axis, this is not necessarily required and we relax this assumption here.

\subsection{Annulus with 5-branes}
With poles on both boundaries the holomorphic functions for the annulus are
\beq\label{eq:genA}
{\cal A}_+(w) = {\cal A}^0_++\alpha_+ w+\sum_{l=1}^LZ^l_+\log\theta_1(w-p_l|\tau),~~~~{\cal A}_-(w) = {\cal A}^0_++\alpha_- w+\sum_{l=1}^LZ^l_+\overline{\log\theta_1(\overline{w}-p_l|\tau)},
\eeq
where
\begin{align}\label{eq:alphaZ}
\alpha_+&=-\overline{\alpha_-}=-\sum_{l=1}^LZ^l_+\partial_{s}\log\theta_1(s-p_l|\tau),\nn\\[2mm]
Z^l_{+}&=-\overline{Z^l_{-}}= \sigma \frac{\prod_{n=1}^L \theta_1 (p_l-s_n|\tau)}{\prod_{n\neq l}\theta_1(p_l-p_n|\tau)} \,\exp\left\{-\frac{2\pi i}{\tau}p_l\Lambda_+-\frac{i\pi}{\tau}\sum_{n=1}^L(p^2_n+p_n)\right\},
\end{align}
and where $s$ is any one zero $s_i$, $\sigma$ is an arbitrary complex constant and 
\beq\label{eq:lambdacond}
\Lambda_+=\sum_{l=1}^L(s_l-p_l),~~~\Lambda_-=\sum_{l=1}^L(\overline{s}_l-p_l),~~~~\Lambda_{\pm}\in \mathbb{Z}\tau.
\eeq
The first thing to note is that clearly, ${\cal A}_{\pm}$ are not invariant under $w\to w+1$, unless $\alpha_{\pm}=0$. This is not necessarily a problem if ${\cal A}_{\pm}(w+1)$ can be expressed as an $SL(2,\mathbb{R})$ of ${\cal A}_{\pm}(w)$ as in \eqref{eq:monodromy} - however this requires $\overline{\alpha_+}=\alpha_-=-\alpha_-=-\overline{\alpha_+}$ and so 
\beq\label{eq:alphacond}
\alpha_{\pm}=0
\eeq
in the absence of 7-branes. In the presence of 7-branes however this is not necessarily the case, and so we shall proceed with $\alpha_{\pm}\neq 0$ for the sake of the next section. One can integrate $\partial_{w}{\cal B}$ as
\begin{align}\label{eq:genB}
{\cal B}(w)&={\cal B}_0+ ({\cal A}^0_+\alpha_--{\cal A}^0_-\alpha_+) w+\sum_{l=1}^L\bigg[({\cal A}^0_++\alpha_+w) Z^l_- \overline{\log\theta_1(\overline{w}-p_l|\tau)}-({\cal A}^0_- Z^l_+-\alpha_-w) \log\theta_1(w-p_l|\tau)\nn\\[2mm]
&-2 \int_1^w dz\big(Z^l_- \alpha_+ \overline{\log\theta_1({w}-p_l|\tau)}-Z^l_+ \alpha_- \log\theta_1(w-p_l|\tau)\big)\bigg],\\[2mm]
&+\sum_{l,l'=1}^L\int_1^wdz \big(Z^{l}_+Z^{l'}_- \log\theta_1(w-p_l|\tau)\overline{\partial_{\overline{w}}\log\theta_1(\overline{w}-p_{l'}|\tau)}-Z^{l'}_+Z^{l}_- \overline{\log\theta_1(\overline{w}-p_l|\tau)}\partial_w\log\theta_1(w-p_{l'}|\tau)\big)\nn
\end{align}
where we take $w_0=1$ as a regular reference point and when performing integration along the boundary, the contour takes infinitesimal semi-circular deviations into the interior of $\Sigma$ to avoid the poles. Given \eqref{eq:genA} and \eqref{eq:genB} we construct the rather involved expression 
\begin{align}
{\cal G}(w)&={\cal G}^0+2({\cal A}^0 \alpha_++\overline{{\cal A}^0} \alpha_-)(w-\overline{w})\\[2mm]
&+\sum_{l=1}^L\bigg[2{\cal A}^0 Z^l_-\big(\overline{\log\theta_1(\overline{w}-p_l|\tau)}-\overline{\log\theta_1(w-p_l|\tau)}\big)+2\overline{{\cal A}^0} Z^l_+\big(\log\theta_1(w-p_l|\tau)-\log\theta_1(\overline{w}-p_l|\tau)\big)\nn\\[2mm]
&+\bigg(\alpha_+Z^l_- \big(\overline{\log\theta_1(\overline{w}-p_l|\tau)}-\overline{\log\theta_1(w-p_l|\tau)}\big)-\alpha_- Z^l_+ \big(\log\theta_1(w-p_l|\tau)-\log\theta_1(\overline{w}-p_l|\tau)\big)\bigg)(w+\overline{w})\nn\\[2mm]
&-2 \bigg(\int^w_1dz\big(\alpha_+Z^l_- \log\theta(z-\overline{p_l}|\tau)-\alpha_-Z^l_+ \log\theta(z-p_l|\tau)\big)+\text{c.c}\bigg)\bigg]\nn\\[2mm]
&+\sum_{l,l'=1}^L\bigg[-Z^l_+ Z^{l'}_-\log\theta_1(w-p_l|\tau)\overline{\log\theta_1(w-p_{l'}|\tau)}+Z^{l'}_+ Z^{l}_-\overline{\log\theta_1(\overline{w}-p_l|\tau)} \log\theta_1(\overline{w}-p_{l'}|\tau)\nn\\[2mm]
&+\bigg(\int_1^wdz\big(Z^l_+ Z^{l'}_-\log\theta_1(z-p_l|\tau)\overline{\partial_{\overline{z}} \log\theta_1(\overline{z}-p_{l'}|\tau)}-Z^{l'}_+ Z^{l}_-\overline{\log\theta_1(\overline{z}-p_l|\tau)}\partial_z \log\theta_1(z-p_{l'}|\tau)\big) +\text{c.c}\bigg)\bigg]\nn,
\end{align}
where ${\cal G}^0=|{\cal A}^0_+|^2-|{\cal A}^0_-|^2+{\cal B}^0+\overline{{\cal B}^0}$, $2{\cal A}^0={\cal A}^0_+-\overline{{\cal A}^0_-}$ and we see that ${\cal G}(w+1)={\cal G}(w)$ indeed requires $\alpha_{\pm}=0$.

Since by definition $\partial_w {\cal A}_{\pm}(w)=-\overline{\partial_{\overline{w}} {\cal A}_{\mp}(\overline{w})}$ and \eqref{eq:peicewiseconstantcond} holds trivially for $w\in \mathbb{R}$ and for $w\in \mathbb{R}+\frac{\tau}{2}$ by the periodicity of $\theta_1(w|\tau)$ under $w\to w+\tau$, it follows that ${\cal G}(w)$ is piece-wise constant on both boundaries. We ensure that ${\cal G}(w)=0$ on each boundary by first fixing ${\cal G}(w)=0$ for $\text{Re}\,w> \text{Re}\,p_l$ for all $l$ and then imposing zero monodromy across the poles - which are the only point at which ${\cal G}$ can jump. For $w=x\in\mathbb{R}$ it is a simple matter to show that
\beq\label{eq:0b}
{\cal G}(x)={\cal G}^0
\eeq
which can be set to zero by tuning ${\cal B}^0$. For $w=x+\frac{\tau}{2}$ things are less trivial, because of the integrals. One can proceed by splitting these as $[w,1]=[x+\frac{\tau}{2},1+\frac{\tau}{2}]\cup [1+\frac{\tau}{2},1]$, and by exploiting the periodicities of $\theta_1(w,\tau)$. One finds that all the dependence on $x$ cancels between integral and non-integral terms, leaving
\begin{align}\label{eq:taub}
\frac{{\cal G}(x+\frac{\tau}{2})}{2\pi i}&=\frac{{\cal G}^0}{2\pi i}+\frac{\tau}{2\pi i}(2{\cal A}^0 \alpha_-+2\overline{{\cal A}^0} \alpha_+)+ 2\sum_{l=1}^L\bigg[Z^l_-({\cal A}^0+\alpha_+) \overline{p_l}+Z^l_+(\overline{{\cal A}^0}-\alpha_-) p_l,\nn\\[2mm]
&-\frac{1}{2\pi i}\bigg(\int_0^{\frac{\tau}{2}}dw\big(Z^l_-\alpha_+\overline{\log\theta_1(\overline{w}-p_l|\tau)}-Z^l_+\alpha_-\log\theta_1(w-p_l|\tau)\big)+\text{c.c}\bigg)\bigg]\\[2mm]
&+\sum_{l,l'=1}^L\bigg[\overline{p_l} Z^{l'}_+ Z^{l}_-\log\theta_1\left(\frac{\tau}{2}+p_{l'}\big|\tau\right)-p_l Z^l_+ Z^{l'}_-\overline{\log\theta_1\left(-\frac{\tau}{2}+p_{l'}\big|\tau\right)}\nn\\[2mm]
&+\frac{1}{2\pi i}\bigg(\int_0^{\frac{\tau}{2}}dw\big(Z^l_+ Z^{l'}_-\log(w-p_l|\tau)\overline{\partial_{\overline{w}} \log(\overline{w}-p_{l'}|\tau)}-Z^{l'}_+ Z^{l}_-\log(w-\overline{p_l}|\tau)\partial_w \log(w-p_{l'}|\tau)\big)+\text{c.c}\bigg)\nn\bigg],
\end{align}
which, while rather complicated, is just one real constraint. Provided \eqref{eq:0b}-\eqref{eq:taub} can be simultaneously set to zero and that \eqref{eq:lambdacond} and \eqref{eq:alphacond} can be likewise solved, we just need to impose zero monodromy as we move around  the poles to have a well-defined solution. As shown in \cite{DHoker:2017mds} this requires the following
\begin{align}\label{eq:deltacalGno7s}
\frac{\Delta_k{\cal G}}{2\pi i \delta_k} &=2 Z^k_-{\cal A}^0_+-2 Z^k_+{\cal A}^0_-+ (Z^k_-\alpha_+-Z^k_+\alpha_-)(p_k+\overline{p_k})+\sum_{l\neq k} Z^{[l,k]}\log|\theta_1(p_k-p_l)|^2\nn\\[2mm]
&- \frac{i\pi}{\tau}\sum_{l=1}^L(p_l-\overline{p_l})\bigg[(1+p_l+ \overline{p_l}-2p_k)Z^k_+ Z^l_-+(1+p_l+ \overline{p_l}- 2 \overline{p_k})Z^l_+ Z^k_-\bigg]
\end{align}
which is a total of $L$ real constraints - reducing to $L-1$ when $\alpha_+=0$ and all poles are on the same boundary, as then $\sum_k \delta_k\Delta_k{\cal G}=0$ automatically. 

Note that all the expressions in this section simplify considerably when we set $\alpha_{\pm}=0$ as is required for periodicity under $w\to w+1$, and when we assume all poles are on the real axis, as we do in the main text.

\subsection{Adding 7-branes}\label{eq:annulusD7appendix}
To include $(p,q)$ 7-branes we must modify \eqref{eq:genA} as
\begin{align}\label{eq:genA7brane}
{\cal A}_{\pm}(w)= {\cal A}^s_{\pm}(w)+\eta_{\pm}\int_1^w dz\,\partial_z{\cal A}^{(0)}(z) f(z),
\end{align}
where the integration contour is taken such that it always remains inside $\Sigma$, avoids crossing the branch cuts associated to the branch points in $f(w)$ and takes a semi-circular path around the poles on the boundary. In \eqref{eq:genA7brane} we use the short hand notation
\begin{align}\label{eq:genA07brane}
{\cal A}^{(0)}(w)&={\cal A}^{(0)}_+(w)-{\cal A}^{(0)}_-(w),\nn\\[2mm]
{\cal A}^s_{+}(w)&=u_{[p,q]}{\cal A}^{(0)}_{+}(w)+v_{[p,q]}{\cal A}^{(0)}_{-}(w),\nn\\[2mm]
{\cal A}^s_{-}(w)&=-\overline{v_{[p,q]}}{\cal A}^{(0)}_{+}(w)+\overline{u_{[p,q]}}{\cal A}^{(0)}_{-}(w),
\end{align}
where $v_{[p,q]},u_{[p,q]},\eta_{\pm}$ are defined in \eqref{eq:uveta}, $f$ in \eqref{f-annulus} and the ${\cal A}^{(0)}_{\pm}$ refer to the holomorphic functions in \eqref{eq:genA}. As such \eqref{eq:genA7brane} contains order $ \mathcal{O}(w)$ terms with coefficients $\alpha_{\pm}$ defined as
\beq
\alpha_{+}=-\overline{\alpha_-}=u_{[p,q]}\alpha^{(0)}_++v_{[p,q]}\alpha^{(0)}_-,
\eeq
where $\alpha^{(0)}_{\pm}$ refers to the $\alpha_{\pm}$ of \eqref{eq:alphaZ}. 

The first thing to address is the periodicity under $w\to w+1$. Clearly ${\cal A}_{\pm}$ are not invariant, indeed they transform as
\beq\label{eq:Ashift}
{\cal A}_{\pm}(w+1)-{\cal A}_{\pm}(w)= \eta_{\pm}\Delta f {\cal A}^{(0)}(w)+\alpha_{\pm}+\eta_{\pm}{\cal T},
\eeq
where
\beq\label{eq:deltaft}
\Delta f= f(w+1)-f(w)=-\frac{i}{2 \tau}\sum_{i=1}^I n_i^2(w_i-\overline{w_i}),~~~~{\cal T}= \int_1^{0} dz {\cal A}^{(0)}(z)\partial_z f(z),
\eeq
which is easiest to see by integrating by parts, expressing the integral between $[1,~w+1]$ as the difference between integrals over $[1,~2]$ and $[w+1,~2]$ and then using the $w\to w+1$ periodicity of ${\cal A}^{(0)}(w),~\partial_w f(w)$.
Clearly the last 2 terms in \eqref{eq:Ashift} are constants, but the first is not - so we must seek to repackage \eqref{eq:Ashift} as an $SL(2,\mathbb{R})$ if we are to achieve periodicity. It is not hard to show that
\begin{align}
{\cal A_+}(w+1)=+ u  {\cal A}_+(w)- v {\cal A}_-(w) + a_+,~~~~
{\cal A_-}(w+1)=-  \overline{v} {\cal A}_+(w)+ \overline{u}{\cal A}_-(w) + a_-,
\end{align}
where 
\beq
u=1+\Delta f\,\eta_+\eta_-,~~~~v=\Delta f\,\eta_+^2,~~~~a_{\pm}=\alpha_{\pm}+{\cal T},~~~~ |u|^2-|v|^2=1,
\eeq
so we know that the (Einstein frame) metric will be invariant provided we impose that $a_-=\overline{a_+}$. This is however insufficient for $w\to w+1$ periodicity, and we also need to impose that the dilaton is invariant under this transformation and that the fluxes potentials transform suitably. One can check using the transformation rules in \eqref{eq:sl2rtrans} that an invariant dilaton requires  fixing $q=0$, which implies we are restricting our considerations to the addition of $(p,0)$ 7-branes - so we can without loss of generality set
\beq
(p,q)=(1,0),~~~~\eta_+=\eta_-=1,~~~~ u_{[1,0]}=1,~~~v_{[1,0]}=0.
\eeq
Given this restriction one finds that the following transformation rules
\beq
e^{\Phi}\to e^{\Phi},~~~ C_0\to C_0-2 i\, \Delta f,~~~~B_2\to B_2,~~~ C_2\to C_2- 2 i\,\Delta f B_2
\eeq
under $w\to w+1$ - which is just a gauge transformation that leaves the fluxes, $H_3=dB_2,~~F_1=dC_0,~~~F_3=dC_2-C_0 H_3$, invariant, and so all the physical fields are now invariant under $w\to w+1$ provided $a_-=\overline{a_+}$. This condition imposes that
\beq\label{eq:alpharules}
\text{Re}\,\alpha_+=0,~~~~ \text{Im}\,\alpha_+=\text{Im} \int^1_0 dw {\cal A}^{(0)}(w)\partial_w f(w).
\eeq 
So ensuring  $w\to w+1$ imposes 2 real constraints in the presence of (what are necessarily) D7 branes, and  we see that unlike the monodromy free case of the previous section, we no longer have $\alpha_+=0$ in general.  At this point we have refined \eqref{eq:genA7brane} to
\begin{align}
{\cal A}_{\pm}(w)= {\cal A}_{\pm}^{(0)}(w)+ \int_{1}^w dz\sum_{l=1}^L\partial_{z}\big(Z^l_{+}\log(z-p_l|\tau)-Z^l_{-}\overline{\log(\overline{z}-p_l|\tau)}\big)f(z)
\end{align}
where $\alpha_{\pm}=\alpha^{(0)}_{\pm}$ appears only in ${\cal A}_{\pm}^{(0)}(w)$, because ${\cal A}^{(0)}(w)$ only depends on these constants in the combination $(\alpha_+-\alpha_-)$, which vanishes by \eqref{eq:alpharules}.

The next thing we should impose is that ${\cal G}(w)$ is continuous as we move around a branch cut. This is ensured if the difference between points on either side of a cut is an $SL(2,\mathbb{R})$ transformation of the form \eqref{eq:monodromy}. As shown in \cite{DHoker:2017zwj} this imposes \eqref{eq: monodomycond}, which more specifically imposes
\beq\label{eq:moncond7annulus}
\text{Re}\bigg[{\cal A}^0+\frac{1}{2}\sum_{l=1}^L\big(Z^l_+\log\theta_1(w_i-p_l|\tau)-Z^l_-\overline{\log\theta_1(\overline{w_i}-p_l|\tau)}\big)\bigg]=0
\eeq
in this case - recall $2{\cal A}^0={\cal A}^0_+-\overline{{\cal A}^0_+}$ is a constant.

Now we must ensure that ${\cal G}(w)$ vanishes when $w\in\mathbb{R}$ and $w\in \mathbb{R}+\frac{\tau}{2}$. The properties of $\partial_w{\cal A}^{(0)}_{\mp}(w)$ and $f$  ensure that it is piece-wise constant, so all we need do is find a region of the boundary where ${\cal G}(w)=0$ and impose that there is no monodromy across the poles - \eqref{eq:moncond7annulus} already ensures ${\cal G}$ is continuous across branch cuts. As in the previous section we choose our region of the real boundary to be $w=x\in\mathbb{R}$ such that $\text{Re}p_l<x\leq 1$ for all $l$. Given the expression for ${\cal G}(w)$ with 7-branes in \eqref{eq:calG7brane} one easily sees that
\beq
{\cal G}(x) = {\cal G}^{(0)}(x)= {\cal G}^0,
\eeq
provided one fixes the integration constants in ${\cal A}^{(0)}_{\pm}(w)$ as
\beq
\overline{{\cal A}^0_{\pm}}=-{\cal A}^0_{\mp},
\eeq
which we are free to do without loss of generality. Thus one can fix ${\cal G}(x)=0$ by tuning ${\cal B}^0$ as in the proceeding section. For the region  $\mathbb{R}+\frac{\tau}{2}$ we choose $w=x+\frac{\tau}{2}$ with $x$ defined as before. In order to make progress here it is helpful to make use of the identity
\beq
{\cal A}^{(0)}(x+\frac{\tau}{2})={\cal A}^{(0)}(x-\frac{\tau}{2})+ i C,~~~ C=2\pi \sum_{l=1}^L \big(Z^l_+p_l- Z^l_-\overline{p_l}\big)\in \mathbb{R},
\eeq
which follows from the quasi periodicity of $\theta_1(w|\tau)$ under $w\to w+\tau$. Additionally it is useful to split the integrals in \eqref{eq:calG7brane} as $[1,~x+\frac{\tau}{2}]=[1,~1+\frac{\tau}{2}]\cup [1+\frac{\tau}{2},~x+\frac{\tau}{2}]$ so that all dependence on $x$ drops out of ${\cal G}$. After further exploiting  the $\theta_1(w|\tau)$ periodicities one finds
\begin{align}\label{eq:genannulusupperbound}
{\cal G}(x+\frac{\tau}{2}) &= {\cal G}^{(0)}(x+\frac{\tau}{2})+2i C\, \Delta f{\cal A}^{(0)}(0)\nn\\[2mm]
&+2\text{Re}\bigg[\int_0^{\frac{\tau}{2}}dz \partial_{z}({\cal A}^{(0)}(z)^2)f(z)\bigg]+2 C\,\text{Im}\bigg[\int_0^{\frac{\tau}{2}}dz\partial_{z}{\cal A}^{(0)}(z)f(z)\bigg]
\end{align}
where care must be taken to avoid any branch cuts that cross $\text{Re}w=0\sim 1$ when performing the integrals.

Last we impose that ${\cal G}$ has zero monodromy across the poles - to do this it is useful to first calculate the discontinuity of its constituent parts. As one integrates along one of the boundaries, one needs to take infinitesimal semi-circular deformations about the pole. Such a semi-circular contour is defined as
\beq
C_k: w= p_k+ \epsilon e^{i \delta_k \theta},~0\leq \theta \leq\pi,~~~\delta_k=1,~p_k\in\mathbb{R},~~~\delta_k=-1,~p_k\in\mathbb{R}+\frac{\tau}{2},
\eeq
where $\epsilon\in \mathbb{R}^+$ is an infinitesimal parameter.
As such the discontinuity of the holomorphic functions across the pole $p_k$ is given by
\beq
\Delta_k{\cal A}={\cal A}_{\pm}(p_k-\epsilon)-{\cal A}_{\pm}(p_k-\epsilon)= \int_{C_k}\partial_w{\cal A}_{\pm}(w)= i\pi\delta_k\big(Z^k_{\pm}+f(p_k)Z^l\big),~~~Z^l=Z^l_+-Z^l_-,
\eeq
where we use that $f(\overline{p_k})=f(p_k)$.
Similarly one can compute the discontinuity of ${\cal B}(w)$ across the pole $p_k$:
\begin{align}\label{eq:deltacalB}
\Delta_k{\cal B}&= \int_{C_k}\partial_w{\cal B}(w)=\Delta_k{\cal B}^{(0)}+\int_{C_k}dw\bigg[{\cal A}^{(0)}(w)\partial_w {\cal A}^{(0)}(w)f(w)-\partial_w {\cal A}^{(0)}(w)\int_1^w dz\partial_z {\cal A}^{(0)}(z)f(z)\bigg]\nn\\[2mm]
&=\Delta_k{\cal B}^{(0)}+ i\pi \delta_k Y^k\bigg(f(p_k){\cal A}^{(0)}(p_k+\epsilon)-\int_1^{p_k+\epsilon}dw\partial_w {\cal A}^{(0)}(w)f(w)\bigg)+...,
\end{align}
where "..." is purely imaginary, so does not feature in what follows.
Putting this together we find  that $\Delta_k(|{\cal A}_+|^2-|{\cal A}_-|^2)$ just contributes the same as $\Delta_k{\cal B}+\text{c.c}$ and after integrating by parts to cancel the mutually divergent terms in the second line of \eqref{eq:deltacalB} one is left with
\begin{align}
\frac{\Delta_k {\cal G}}{2\pi i \delta_k}&=\frac{\Delta_k {\cal G}^{(0)}}{2\pi i \delta_k}+Z^k\bigg[2\big({\cal A}^0+\overline{{\cal A}^0}\big)\big(f(p_k)-f(1)\big)+2 A^{(0)}(1)f(1)\nn\\[2mm]
&+\bigg(\int_1^{p_k}dw\sum_{l=1}^L\big(Z^l_+\log\theta_1(w-p_l|\tau)-Z^l_-\overline{\log\theta_1(\overline{w}-p_l|\tau)}\big)\partial_w f(w)-\text{c.c}\bigg)\bigg],
\end{align}
where as before, $\Delta_k {\cal G}^{(0)}$ is given in \eqref{eq:deltacalGno7s}, and the integration contour is taken such that it crosses no branch cuts. 

\section{Orientifolds and non-Abelian T-duality}
\label{worldsheet}

In this Appendix we use the string sigma model to derive the transformation of worldsheet parity reversal under non-Abelian T-duality. This will identify the way in which the 
O8 orientifold fixed plane of the original Brandhuber-Oz solution is transformed. 

The most general sigma model describing a string propagating on a NS-NS background invariant under a non-isotropic\footnote{Namely, an isometry group acting without fixed points.} non-Abelian isometry group $G$ is 
\begin{eqnarray}
S[g,x]&=&\int d\sigma_+ d\sigma_- \Bigl[E_{ab}(x)(\partial_+ g g^{-1})^a(\partial_- g g^{-1})^b+F^{R}_{a\alpha} (\partial_+ g g^{-1})^a \partial_- x^\alpha+\nonumber\\
&&+F^{L}_{\alpha a}(x)\partial_+ x^\alpha (\partial_- g g^{-1})^a
+F_{\alpha\beta}(x)\partial_+ x^\alpha\partial_- x^\beta\Bigr] ,
\end{eqnarray}
where $g\in G$, a compact Lie group, and $\partial_\pm g g^{-1}=(\partial_\pm g g^{-1})^a T_a$, with $T_a$ the generators of the corresponding Lie algebra\footnote{Normalised such that ${\rm Tr}(T_a T_b)=\delta_{ab}$.}. This model is invariant under $g\rightarrow g h$, with $h\in G$. 

The non-Abelian T-dual of this sigma model with respect to the full isometry group $G$ can be constructed using the gauging procedure, as in 
\cite{delaOssa:1992vci}, or the canonical transformation approach, as in \cite{Alvarez:1994wj,Lozano:1995jx}. The last approach is particularly useful for studying the non-local map that underlies non-Abelian T-duality, and generalises the well-known transformation
\begin{equation}
\label{circle}
\partial_+\phi=-\frac{1}{R^2}\partial_+ {\tilde \phi}\, ,  \qquad \partial_-\phi=\frac{1}{R^2}\partial_- {\tilde \phi},
\end{equation}
for a string propagating in a circle of radius $R$, to the non-Abelian case. 
This mapping was derived in \cite{Borlaf:1996na}, using the canonical transformation approach \cite{Lozano:1995jx}.
In the particular case in which $F^L_{\alpha b}=F^R_{b\alpha}=0$, and thus the string sigma model is invariant under $G_L\times G_R$, as the BO solution, it simplifies to
\begin{equation}
\label{PR}
(\partial_+ g g^{-1})^a=-M_{ba}\,\partial_+\chi\, , \qquad (\partial_- g g^{-1})^a=M_{ab}\,\partial_-\chi ,
\end{equation}
where $M\equiv (E+{\rm ad}\chi)^{-1}$ and $\chi=\chi^a T_a$ are the new variables that parameterise the propagation of the string in the non-Abelian T-dual background. These variables live, by construction, in the Lie algebra associated to the $G$ isometry group. 

Using the mapping (\ref{PR}) we can now identify the way worldsheet parity reversal transforms under non-Abelian T-duality. Worldsheet parity reversal interchanges the right and left moving sectors of the original sigma model. Using the mapping (\ref{PR}), this interchanges these sectors in the dual sigma model, and inverts $\chi \rightarrow -\chi$. Therefore, 
\begin{equation}
\label{chiomega}
\Omega \rightarrow I_\chi \Omega.
\end{equation}
This generalises the well-known mapping $\Omega\rightarrow I_{\tilde \phi}\Omega$ under Abelian T-duality, implied by (\ref{circle}), to the non-Abelian case.
The transformation (\ref{chiomega})  is a general result, valid for any string sigma model invariant under $G_L\times G_R$, dualised with respect to one of these isometry groups.

Particularising this result to the non-Abelian T-dual of the BO solution, we find that $I_\theta\Omega\rightarrow I_\theta I_\chi \Omega$ after the duality. This transformation thus maps the 
O8 orientifold fixed plane of the BO solution, located at $\theta=\pi/2$, onto a O5 fixed plane located at $\theta=\pi/2$ and at the origin of $\mathbb{R}^3$. In section \ref{NATD} the $\mathbb{R}^3$ space was parameterised in spherical coordinates, such that the $SU(2)$ remaining symmetry was manifest. In these coordinates the O5 fixed plane is thus located at $r=0$.

\section{Spin-2 excitations}
\label{spin2}

Linear fluctuations of the gravity background around the $AdS_6$ vacuum are expected to be dual to operators that belong to $F(4)$ superconformal multiplets \cite{Buican:2016hpb,Cordova:2016emh}. In reference \cite{Gutperle:2018wuk} the equations governing spin-2 fluctuations of the Type IIB $AdS_6\times S^2 \times \Sigma$ solutions of DGKU/DGU were derived. These are known to depend only on the background metric due to the universality of the spin-2 mass operator \cite{Bachas:2011xa}.
 
In particular, a transverse traceless perturbation for the $AdS_6$ factor of the metric in \eqref{eq:metric_DHoker} was considered,
\begin{equation}
ds^2 = \lambda_6^2 \left(ds^2(AdS_6) +h_{\mu\nu}(x,y)dx^\mu dx^\nu\right) + \lambda_2^2 \,ds^2(S^2)+ 4\tilde{\rho}^2dwd\bar w \,,
\end{equation} 
where $x$ and $y$ denote collectively the $AdS_6$ and internal space directions, respectively. The behaviour of $h_{\mu\nu}$ is given by \cite{Bachas:2011xa}:
\begin{equation}\label{eq:Laplace_gravitons}
\frac{1}{\sqrt{|g|}}\partial_A \left(\sqrt{|g|}\, g^{AB}\partial_B h_{\mu\nu} \right)=0 \,,
\end{equation}
where $g$ is the 10d metric and $A,B=0,1,\ldots,9$. Assuming the mode expansion
\begin{equation}
h_{\mu\nu}(x,y)= h_{\mu\nu}^{[tt]}(x)\, \Psi(y) \,,
\end{equation}
equation \eqref{eq:Laplace_gravitons} decomposes into two separated equations for modes of mass $M$ on the external and internal spaces:
\begin{equation}
\square_{AdS_6}h^{[tt]}_{\mu\nu}= ( M^2 -2) h^{[tt]}_{\mu\nu} \,
\end{equation} 
for the $AdS_6$ directions, and 
\begin{equation}\label{eq:gravitons_internal}
-{1\over \lambda_6^4 \lambda_2^2 \tilde{\rho}^2} \partial_a\big(\lambda_6^6\lambda_2^2  \eta^{ab} \partial_b \Psi\big )- {\lambda_6^2 \over \lambda_2^2} \nabla_{S^2}^2 \Psi = M^2 \Psi \,,
\end{equation}
where $a,b= w, \bar w$ and $\eta^{w\bar w}=\eta^{\bar{w} w}=1/2$, for the $\Psi$ modes on the $S^2$ and Riemann surface $\Sigma$. A further expansion in spherical harmonics on the $S^2$,
\begin{equation}
\Psi(y)= \phi_{\ell m}(w,\bar{w})\,Y_{\ell m}(S^2) \,,
\end{equation}
followed by a field redefinition $\phi_{\ell m}= \mathcal{G}^\ell\,\chi_{\ell m}$, yields to an 
equation for the $\chi_{\ell m}(w,\bar w)$ modes on $\Sigma$, involving only the $\mathcal{G}$ and $\kappa^2$ fields:
\begin{equation}\label{eq:gravitons_sigma}
\partial_a \left(\mathcal{G}^{2\ell+2}\,\partial^a\chi_{\ell m}\right)+\frac{1}{6}\Big(M^2-3\ell(3\ell+5)\Big)\kappa^2\,\mathcal{G}^{2\ell+1}\chi_{\ell m}=0 \,.
\end{equation}

We now study this equation for the Abelian and non-Abelian T-dual geometries considered in this paper.

\subsection{Abelian T-dual case}

The Abelian T-dual case was already studied in  \cite{Gutperle:2018wuk}. Here we take the same choice of holomorphic coordinate $w$ to compare with their results, $w=x+iy$, where $x=(\cos\theta)^{2/3}$ and $y= \beta \psi$, with $\beta=4 m^{1/3}/(3 L^2)$, as in eq.~\eqref{omega}. The corresponding $\mathcal{G}$ and $\kappa^2$ functions can be readily deduced from \eqref{eq:AdS6_sol_ATD}:
\begin{equation}\label{eq:G_k_ATD}
\mathcal{G} = \frac{81}{128}\,L^6(1-x^3) \,, \qquad \kappa^2 = \frac{243}{256}\,L^6 x \,.
\end{equation}
Inserting these in eq.~\eqref{eq:gravitons_sigma}, we get for $\chi_{\ell m}(x,y)$ and $x<1$:
\begin{equation}\label{eq:chi_ATD}
(1-x^3) (\partial_x^2+ \partial_y^2)\chi_{\ell m} - 6(\ell+1)x^2 \partial_x\chi_{\ell m} +\Big(M^2-3\ell(3\ell+5)\Big) x\,\chi_{\ell m} =0\,.
\end{equation}
Following the discussion in \cite{Gutperle:2018wuk}, we take advantage of the $U(1)$ isometry of the Abelian T-dual solution along $y=\beta\psi$ to expand $\chi_{\ell m}$ as
\begin{equation}\label{eq:chi_decomp}
\chi_{\ell m}(x,y)= f(x)\,e^{iky} ,
\end{equation}
for a real constant $k$. Back to eq.~\eqref{eq:chi_ATD}, this decomposition yields an ODE for $f(x)$ which determines the spin-2 fluctuations of the Abelian T-dual solution:
\begin{equation}\label{eq:f_ATD}
(1-x^3) \left( f''(x) -k^2f(x) \right) - 6(\ell+1)x^2 f'(x) +\Big(M^2-3\ell(3\ell+5)\Big) x f(x) =0\,.
\end{equation}
This equation reduces for $k=0$ to a hypergeometric differential equation.

\subsection{Non-Abelian T-dual case}

We now turn to the non-Abelian T-dual solution presented in \eqref{eq:AdS6_sol_NATD}. We take the  same choice $w=x+iy$ for the holomorphic coordinate with $y= \beta r$ in this case. Again, we read-off $\mathcal{G}$ and $\kappa^2$ from the metric factors,
\begin{equation}\label{eq:G_k_NATD}
\mathcal{G} = \frac{81}{128}\,L^6r(1-x^3) \,, \qquad \kappa^2 = \frac{243}{256}\,L^6 r\,x \,.
\end{equation}
Note that they coincide with the expressions for the Abelian T-dual solution, multiplied by an $r$ factor.
Substituting in equation \eqref{eq:gravitons_sigma} we then get:
\begin{align}\label{eq:chi_NATD}
(1-x^3)y (\partial_x^2+ \partial_y^2)\chi_{\ell m} - 6(\ell+1)x^2y\, \partial_x\chi_{\ell m} - 2(\ell+1)(1-x^3)y\,\partial_y\chi_{\ell m}&\\[2mm] +\Big(M^2-3\ell(3\ell+5)\Big) xy\,\chi_{\ell m} &=0\,. \nonumber
\end{align}
Note that this is the same equation \eqref{eq:chi_ATD} derived for the Abelian T-dual solution, multiplied by a $y$ factor, except for the term in $\partial_y\chi_{\ell m}$, not present in the Abelian T-dual solution due to its $y$-isometry.
This implies that any solution $\chi_{\ell m}$ of \eqref{eq:chi_ATD} which is also independent of $y$ will be automatically a solution of \eqref{eq:chi_NATD} for $y>0$. We can therefore use any $f(x)$ solving \eqref{eq:f_ATD} for $k=0$ to find fluctuations in the non-Abelian T-dual background.
On the other hand, given that the non-Abelian T-dual background is no longer isometric in $y$, it is not possible to use the decomposition Ansatz \eqref{eq:chi_decomp}  to find fluctuations depending on both directions of the Riemann surface.


\subsection{Some classes of fluctuations}

In \cite{Gutperle:2018wuk} two classes of universal fluctuations were considered common to all DGKU/DGU geometries. Relying on the fact that $\mathcal{G}$ and $\kappa^2$ are non-negatively defined, it was shown that all solutions must satisfy the bound $M^2 \geq 3\ell(3\ell+5)$.  The simplest class of solutions are those saturating this bound, with constant $\chi_{\ell m}$. These were referred as minimal solutions in \cite{Gutperle:2018wuk}. In the Abelian case this is an obvious solution of the hypergeometric equation that is obtained setting $k=0$ in \eqref{eq:f_ATD}. As it is independent of $y$, it is also a solution of \eqref{eq:chi_NATD} for the non-Abelian T-dual case.

The second class of universal solutions considered in \cite{Gutperle:2018wuk}, tagged non-minimal, used  the decomposition $\chi_{\ell m}= \mathcal{A}_+ - \bar{\mathcal{A}}_-$ for $M^2 = (3\ell+1)(3\ell+6)$. Both equation \eqref{eq:chi_ATD}, for the Abelian case, and  \eqref{eq:chi_NATD}, for the non-Abelian one, admit solutions within this class, as expected. 
In order to write them down, we just need to adapt the expressions for $\mathcal{A}_\pm$ given in eqs. \eqref{eq:holo_Ansatz_ATD_new} and \eqref{ApmNA} to our new choice $w=x+i y$ with $x=(\cos\theta)^{2/3}$ and $y=\psi,r$ for the holomorphic coordinate,
\begin{align}
\mathcal{A}_\pm^{\textrm{ATD}} &= -\frac{3m}{8\b}w^2 \pm \frac{3}{4\b}w \,, \nonumber\\
\mathcal{A}_+^{\textrm{NATD}} &= i\frac{m}{8\b^3}w^3 + \frac{3}{4\b}w  \,, \qquad
\mathcal{A}_-^{\textrm{NATD}} = i\frac{m}{8\b^3}w^3 - \frac{3}{4\b}w + i\frac{m}{2\b^3} \,.
\end{align}
This yields:
\begin{align}\label{eq:chi_non-minimal}
\chi_{\ell m}^{\textrm{ATD}} &= \frac{3}{2\b}x - i \frac{3m}{2\b^2}xy \,, \nonumber\\[2mm]
\chi_{\ell m}^{\textrm{NATD}} &= \frac{3}{2\b}x + i \frac{m}{4\b^3}(x^3-3xy^2+2) \,.
\end{align}
One can see that both the real and imaginary parts of these expressions are solutions on their own right, admitting arbitrary overall constants.

Another class of solutions common to the Abelian and non-Abelian T-dual backgrounds is given by $\chi_{\ell m}=f(x)$, with $f(x)$ satisfying the hypergeometric differential equation \eqref{eq:f_ATD} for $k=0$ and any mass $M^2 \geq 3\ell(3\ell+5)$, 
\begin{align}\label{eq:f_hyper}
f(x) &= C \; \phantom{}_2F_1\left(\frac{5}{6}+\ell-\frac{1}{6}\sqrt{4 M^2+25}\,,\,\frac{5}{6}+ \ell + \frac{1}{6}\sqrt{4 M^2+25}\,;\,\frac{2}{3}\,;\,x^3\right) \nonumber\\[2mm] 
& +\tilde C x \; \phantom{}_2F_1\left(\frac{7}{6}+\ell-\frac{1}{6} \sqrt{4 M^2+25} \,,\, \frac{7}{6}+\ell+\frac{1}{6} \sqrt{4 M^2+25} \,;\, \frac{4}{3} \,;\, x^3\right) \,,
\end{align}
where $C,\tilde C \in \mathbb{R}$. These solutions overlap with the universal minimal solutions, that can be recovered setting $M^2 = 3\ell(3\ell+5)$, and with the non-mimimal solutions $f(x)= \tilde C\, x$ when $M^2 = (3\ell+1)(3\ell+6)$.


It is worth mentioning that, apart from the non-minimal solution in \eqref{eq:chi_non-minimal}, we can find modes with a simple $y$-dependence allowing $\chi_{\ell m}(x,y)=f(x)\,g(y)$. Substituting in  \eqref{eq:chi_NATD} this gives,
\begin{align}\label{eq:chi_separated}
y\,g(y) \big( (1-x^3)f''(x) - 6(\ell+1)x^2f'(x) \big) + \big(M^2-3\ell(3\ell+5)\big) x\, y f(x)\,g(y) & \nonumber \\[2mm] 
+ (1-x^3)f(x)\left( y\,g''(y)+2(\ell+1)g'(y) \right) &=0 \,.
\end{align}
In the minimal case $M^2=3\ell(3\ell+5)$, we can separate variables assuming $f,g>0$ and solve this equation making the first and last parentheses vanish separately. This is achieved by
\begin{equation}\label{eq:g_NATD}
g(y) = c_1 \,y^{-2\ell -1} + c_2 \,, \qquad c_1,c_2\in \mathbb{R} \,,
\end{equation}
while the required $f(x)$ is just \eqref{eq:f_hyper} with minimal graviton mass.

As a final remark, it was already observed in \cite{Gutperle:2018wuk} that the requirement of having finite fluctuations on $\Sigma$ and its boundaries has a difficult implementation in the Abelian T-dual solution. This is also the case for the non-Abelian T-dual one, for in both backgrounds the full 10d geometry is singular at the boundaries of $\Sigma$, obscuring the interpretation of a $\phi_{\ell m}= \mathcal{G}^\ell\,\chi_{\ell m}$ mode becoming infinite there. This happens for instance in the last example given in \eqref{eq:g_NATD}.

\bibliographystyle{JHEP}
\bibliography{AdS6_bibliography.bib}

\end{document}